\documentclass[fleqn,usenatbib]{mnras}
\usepackage{fix-cm}

\usepackage{newtxtext,newtxmath}

\usepackage[T1]{fontenc}
\usepackage[normalem]{ulem}
\DeclareRobustCommand{\VAN}[3]{#2}
\let\VANthebibliography\thebibliography
\def\thebibliography{\DeclareRobustCommand{\VAN}[3]{##3}\VANthebibliography}

\usepackage{graphicx}	%
\usepackage{amsmath}	%

\usepackage{hyperref}
\usepackage{array}
\usepackage{booktabs}
\usepackage{footmisc} %
\usepackage{color}
\usepackage[table]{xcolor}
\usepackage{placeins}
\usepackage[normalem]{ulem}

\title[Evolution of Binary Systems with a Black Hole Component]{Evolution of Binary Systems with a Black Hole Component: Low- and Intermediate-Mass Donor Systems}

\author[K. Yakut et al.]{
Kadri Yakut,$^{1,2}$\thanks{E-mail: kadri.yakut@ege.edu.tr (KY)}
Belinda Kalomeni,$^{1}$
Saul A. Rappaport,$^{3}$
Lorne A. Nelson,$^{4}$
Christopher A. Tout,$^{2}$
\newauthor
Shunyi Lan,$^{5}$
Jean-Marie Hameury,$^{6}$
and Hai-Liang Chen$^{5}$
\\
$^{1}$Department of Astronomy and Space Sciences, Faculty of Science,
Ege University, 35100, {\.I}zmir, Türkiye\\
$^{2}$Institute of Astronomy, University of Cambridge, Madingley Road,
Cambridge CB3 0HA, United Kingdom\\
$^{3}$Department of Physics and Kavli Institute for Astrophysics and
Space Research, Massachusetts Institute of Technology, Cambridge,
MA 02139, USA\\
$^{4}$Department of Physics and Astronomy, Bishop's University,
2600 College St, Sherbrooke, QC J1M 1Z7, Canada\\
$^{5}$International Centre of Supernovae (ICESUN), Yunnan Key Laboratory
of Supernova Research,\\
\hspace{0.4em}Yunnan Observatories, Chinese Academy of Sciences,
Kunming 650216, China\\
$^{6}$Université de Strasbourg, CNRS, Observatoire astronomique de Strasbourg,
UMR 7550, 67000 Strasbourg, France\\
}
\date{Accepted XXX. Received YYY; in original form ZZZ}

\pubyear{\the\year{}}

\begin{document}
\label{firstpage}
\pagerange{\pageref{firstpage}--\pageref{lastpage}}
\maketitle

\begin{abstract}
We compare secular binary-evolution calculations with the observed properties of Galactic black-hole low-mass X-ray binaries and examine how X-ray irradiation modifies their mass-transfer histories. We analyse about 79,000 irradiation-free MESA sequences from a regular grid (GRID) and a binary population synthesis (BPS) seeded sample, plus about 4,400 irradiated tracks. Comparisons with dynamically constrained black hole binaries use orbital period, donor mass, mass ratio and effective temperature. Both irradiation-free samples produce converging tracks overlapping observed short-period, low-mass-donor systems. Initially more massive donors populate a high-transfer-rate, long-period ascending branch, while transitional systems evolve towards lower mass ratios, cooler temperatures and shorter periods under angular-momentum losses. Relative occupancy differs between samples because of their initial sampling. The irradiated sample shows similar converging and expanding branches. Three illustrative irradiated histories exhibit mass-transfer cycles spanning several orders of magnitude, accompanied by relatively small steps in donor mass and orbital period. Their non-irradiated counterparts follow similar secular trends and timescales. Conditional on the formation of incipient black-hole binaries with low-mass companions under the adopted initial distributions, about 2,600 irradiation-free tracks reach a presently unconfirmed, potentially long-lived class with \(M_2<0.1\,M_\odot\) and \(10\,{\rm min}\leq P_{\rm orb}\leq3\,{\rm h}\). Low transfer rates imply donor mass-loss timescales reaching Gyr scales. A separate irradiation-free grid produces compact, detached CO-core remnants with masses of \(0.501\)--\(0.798\,M_\odot\) and orbital periods of about 230--1300~d. Some retain helium burning and have not settled onto white-dwarf cooling sequences. These results describe evolutionary reachability, not Galactic source abundances or detection yields.
\end{abstract}

\begin{keywords}
X-rays: binaries --- binaries: close --- stars: black holes ---
stars: evolution --- stars: low-mass --- gravitational waves
\end{keywords}

\section{Introduction}\label{sec:intro}

Black-hole low-mass X-ray binaries (BH--LMXBs) provide a unique laboratory to study compact-object formation, long-term mass transfer and angular-momentum loss in interacting binaries.  The dynamically confirmed Galactic sample is dominated by systems with Roche-lobe filling, low-mass donors that are typically discovered as transients, making observational selection effects inseparable from any attempt to infer the underlying population \citep[e.g.][]{Kalogera1999,Jonker2004,Li2015}. Beyond their role as accretion-powered X-ray sources, BH--LMXBs place important constraints on how stellar-mass black holes form in binaries, through mass ejection and natal kicks and how the resulting close binaries subsequently evolve through prolonged phases of mass transfer.

Despite decades of theoretical and observational progress, the standard formation picture still faces a central bottleneck: it is difficult to produce a close BH$+$main-sequence (or subgiant) binary following the common-envelope (CE) phase initiated by the massive black-hole progenitor.  In the classical scenario, the post-CE separation must be small enough for Roche-lobe overflow (RLOF) to occur within a Hubble time yet large enough to avoid merging during the spiral-in.  For low-mass companions this requirement is particularly severe because the available orbital energy is often insufficient to unbind the tightly bound envelope of the massive progenitor under standard $\alpha$--$\lambda$ prescriptions \citep[e.g.][]{Kalogera1999,Podsiadlowski2003,Justham2006}.  This CE survival problem remains one of the most persistent uncertainties in connecting primordial binaries to the observed population of short-period BH--LMXBs.

A widely discussed route around the CE bottleneck is that many present-day BH--LMXBs may descend from binaries with initially intermediate-mass donors (IMXBs). These can more readily survive the CE phase and later be stripped to low masses through mass transfer \citep[e.g.][]{Podsiadlowski2003,Li2015}.  This evolutionary connection was explored in early grid-based studies with interpolated detailed binary sequences \citep{Rappaport2005} and remains an attractive framework to reconcile CE energetics with the existence of compact BH binaries.  However IMXB channels introduce additional problems.  In particular, models that successfully produce close BH binaries by intermediate-mass progenitors often predict donors that are systematically hotter than the observed late-type (K/M) companions.  This effective-temperature mismatch has been emphasized in the context of irradiation-assisted and magnetic-braking driven evolution \citep{Justham2006} and persists in modern calculations that employ updated angular-momentum loss prescriptions and detailed stellar structure \citep{Fan2024}.  Establishing when evolved, chemically processed donors can reproduce the observed donor temperatures therefore remains a key open problem.

A complementary study by \citet{DengLi2024} combined binary-population
synthesis with detailed \textsc{MESA}
Modules for Experiments in Stellar
Astrophysics; \citealt{Paxton2011})
calculations to examine the sensitivity of BH--LMXB populations to four magnetic-braking
prescriptions and four black-hole formation prescriptions.
Our study is based on a set of detailed models
approximately an order of magnitude larger than that of \cite{DengLi2024} and focuses on
different aspects of the evolution.
They found that two of the four magnetic-braking prescriptions, namely
those proposed by \citet{Rappaport1983} and
\citet{ReinersMohanty2012}, generally provide better agreement with the
observed black-hole masses, donor properties, orbital periods, and mean
accretion rates than the other two prescriptions. Their emphasis on
formation and magnetic-braking uncertainties is complementary to our
focus on evolutionary morphology, relative dwell times, and
donor-irradiation feedback.

Individual systems also constrain black-hole birth and subsequent
binary evolution. The survival of the wide
(\(\gtrsim3500\,\mathrm{au}\)) tertiary around V404~Cyg limits the mass
loss and natal kick that could have accompanied black-hole formation and
shows that higher-order multiplicity may contribute to BH--LMXB formation
\citep{Burdge2024,Shariat2025}. More generally, BH X-ray-binary kinematics
permit a range of natal kicks \citep[e.g.][]{Nelemans1999}, while
accretion-driven spin-up can cause present-day black-hole spins to retain
information about the integrated mass-transfer history
\citep{Fragos2015}.

The \textit{Gaia} mission has opened a complementary discovery
channel by identifying dormant black-hole companions in wide binaries
with little or no accretion-powered emission
\citep{ElBadry2023a,GaiaBH3,MuellerHorn2026}. These discoveries demonstrate
that black holes can be found beyond X-ray selected samples and establish
high-precision astrometry, together with spectroscopy, as a practical
route to discovering otherwise hidden black-hole binaries. Continued
astrometric searches should therefore enlarge the known sample and provide
new tests of black-hole binary formation.

Our aim is not to resolve common-envelope ejection. Instead, starting
from already formed black-hole--donor binaries, we examine which regions
of post-formation parameter space overlap the observed BH--LMXB
population under the adopted secular-evolution and irradiation
prescriptions.

To this end, we construct two complementary initial binary sets:
(i) a regular grid of black-hole--main-sequence binaries spanning
black-hole mass, donor mass, and orbital period, and
(ii) an auxiliary BPS-seeded catalogue that provides a complementary,
irregular sampling of already formed black-hole--donor binaries.
Systems from both initial sets are subsequently evolved with
\textsc{MESA}. The detailed \textsc{MESA} sequences do not calculate
black-hole formation or common-envelope ejection, and no
binary-formation weights are assigned to either initial set.
The model--data comparison allows us to identify evolutionary
pathways that overlap the observed BH--LMXB distribution and to quantify
systematic mismatches.

Gravitational-wave observations provide an independent probe of
black-hole formation and evolution
\citep[for a review, see][]{Barack2019}. As a secondary outcome, we
examine whether the stripped-donor and compact-remnant candidates
produced by these calculations enter the low-frequency
gravitational-wave regime. We evaluate the gravitational-wave characteristic strain, \(h_c\), at every valid stored state of the irradiation-free GRID and BPS tracks, rather than selecting only endpoints or minimum-period states.

\section{Observed BH--LMXB sample}
\label{sec:observedBHs}

The observational basis of this study is a curated sample of Galactic black-hole low-mass X-ray binaries (BH--LMXBs) with dynamically inferred system parameters.  These systems are predominantly Roche-lobe filling binaries discovered as transient X-ray sources, reflecting strong observational selection effects that favour short-lived outbursts and systems with favourable viewing geometries. While the total number of known X-ray binaries hosting compact objects is substantially larger, only a limited subset currently provides sufficiently robust constraints on black-hole mass, donor properties and orbital parameters to allow meaningful comparison with detailed binary-evolution models.

Our working sample, summarised in Table~\ref{tab:BHList}, consists of systems for which orbital periods are securely measured and at least partial dynamical constraints on the black-hole mass are available.  In most cases, black-hole masses are derived from the optical mass function combined with estimates of the mass ratio and orbital inclination, obtained through a combination of radial-velocity measurements, ellipsoidal light-curve modelling and, in some systems, constraints from X-ray or radio observations.  Donor-star effective temperatures are typically inferred from spectral classification during quiescence, where contamination from the accretion disc is minimised, although residual systematic uncertainties remain unavoidable. 
We adopt these published constraints rather than re-modelling the disc
luminosity or optical light curve of each observed binary. Residual disc
veiling is therefore included among the observational systematic
uncertainties affecting the comparison.

The inferred donor masses and mass ratios in BH--LMXBs are subject to larger uncertainties than orbital periods or mass functions.  Inclination estimates often dominate the error in black-hole mass determinations, while donor masses depend sensitively on assumptions about spectral type, veiling by the accretion flow and the evolutionary state of the donor star.  As a result, individual systems may admit a range of plausible $(M_{\rm BH}, M_{\rm donor})$ combinations. 
In this work, the observed parameters are treated as constraints with finite uncertainties and the emphasis is placed on the overall distribution of systems in observable parameter space, not on detailed fits to individual binaries.
Throughout this paper, \(M_2\) denotes the donor mass and
\(M_{\rm BH}\) denotes the black-hole mass.

It is important to emphasise that the observed BH--LMXB sample is not a
statistically complete representation of the underlying Galactic population.
Within our calculated stripped-donor BH sequences, short-period
systems with very low-mass donors generally reach low secular mass-transfer
rates. Under the disc-instability interpretation, such rates may correspond
to long quiescent intervals and infrequent outbursts, although the present
calculations do not model these disc cycles. Low donor mass does not by
itself imply a low transfer rate; other compact-binary classes, including
AM~CVn systems, can occupy substantially different mass-transfer regimes.
At sufficiently low secular mass-transfer rates, a disc with a truncated
inner region may instead remain stable on the cold branch
of the thermal--viscous equilibrium curve
on which the disc lies during quiescence, producing a faint
persistent source that would be extremely difficult to identify dynamically.
Very faint X-ray binaries, including ultracompact transient candidates,
provide a related observational context \citep{HameuryLasota2016}.

As a result, such systems may spend the vast majority of their lifetimes below
typical X-ray survey detection thresholds, while dynamical mass measurements
often require a favourable combination of inclination, optical brightness and
quiescent observability.
Consequently, the sample is shaped by strong selection effects that must be
borne in mind when comparing with theoretical populations.
In the following sections, we use this observational sample as a reference
framework against which we map the outcomes of detailed binary-evolution
calculations in the $P_{\rm orb}$--$M_{\rm donor}$,
$q$--$P_{\rm orb}$ and $P_{\rm orb}$--$\log T_{\rm eff}$ planes,
with the aim of identifying which evolutionary channels are consistent with
the data and where significant tensions remain.

\begin{table*}
\centering
\caption{Observed binary parameters adopted for comparison with the
evolutionary calculations. The first block contains 20 Galactic black-hole X-ray binaries with direct or partial dynamical constraints. The second contains
six indirect or quasi-dynamical systems without a donor absorption-line
radial-velocity solution. Numerical superscripts attached to the system names
denote the evidence class defined below. Reference codes combine the
initial of the first author's surname with the two-digit publication
year. Ranges and one-sided limits are retained rather than converted to
symmetric uncertainties.}
\label{tab:BHList}
\scriptsize
\setlength{\tabcolsep}{3.4pt}
\resizebox{\textwidth}{!}{%
\begin{tabular}{r l l l c c c c c c c c}
\hline
ID & System & Alias & S. type &
$P_{\rm orb}/${\rm h} & $M_{\rm BH}/$$M_\odot$ & $M_2/$$M_\odot$ & $q$ & $i/{}^\circ$ & $T_{\rm eff}/$K & $d/{\rm kpc}$ & Ref. \\
\hline
\multicolumn{12}{l}{\textit{A. Direct and partial dynamical constraints}} \\
\hline
1 & XTE J1118+480\textsuperscript{1} & KV UMa & K7--M1 V & $4.0784088\pm0.0000005$ & $6.9$--$8.2$ & $0.18\pm0.06^{\rm D}$ & $0.024\pm0.009$ & $68$--$79$ & $(\sim4000)^{\rm M}$ & $1.72\pm0.10$ & T04, K13 \\
2 & GRO J0422+32\textsuperscript{1} & V518 Per & M4--M5 V & $5.076528\pm0.000011$ & $6.5\pm2.9$ & $0.47\pm0.21^{\rm D}$ & $0.11^{+0.05}_{-0.02}$ & $33$--$49$ & $(\sim3200)^{\rm M}$ & $2.49\pm0.30$ & C24, W00 \\
3 & XTE J1859+226\textsuperscript{1} & V406 Vul & K5--K7 V & $6.624\pm0.072$ & $7.8\pm1.9$ & $0.55\pm0.16^{\rm D}$ & $0.07\pm0.01$ & $66.6\pm4.3$ & $(\sim4100)^{\rm M}$ & $12.5\pm1.5$ & Y22 \\
4 & GRS 1009-45\textsuperscript{2} & MM Vel & K7--M0 V & $6.8449\pm0.0003$ & $\geq4.4$ & $(0.50$--$0.65)^{\rm C}$ & $0.055\pm0.010$ & $37$--$80$ & -- & $3.8\pm0.3$ & F99, S96 \\
5 & XTE J1650-500\textsuperscript{2} & -- & $\sim$K4 V & $7.69\pm0.02$ & $\leq7.3$ & -- & -- & $>47$ & -- & $2.6\pm0.7$ & O04 \\
6 & 1A 0620-00\textsuperscript{1} & V616 Mon & K2--K7 V & $7.752340\pm0.000002$ & $5.86\pm0.24$ & $0.40\pm0.045^{\rm D}$ & $0.060\pm0.004$ & $54.1\pm1.1$ & $(\sim4600)^{\rm M}$ & $1.06\pm0.10$ & V17, C10 \\
7 & GS 2000+25\textsuperscript{1} & QZ Vul & K3--K7 V & $8.258095\pm0.000005$ & $5.5$--$8.8$ & $(0.16$--$0.47)^{\rm C}$ & $0.04\pm0.01$ & $54$--$60$ & -- & $2.7\pm0.7$ & C91, C95, I04, H96 \\
8 & MAXI J1305-704\textsuperscript{1} & -- & K3--K5 & $9.5\pm0.1$ & $8.9^{+1.6}_{-1.0}$ & $0.43\pm0.16^{\rm D}$ & $0.05\pm0.02$ & $72^{+5}_{-8}$ & $4610^{+130}_{-160}{}^{\rm S}$ & $7.5^{+1.8}_{-1.4}$ & M21 \\
9 & GS 1124-684\textsuperscript{1} & GU Mus & K3--K5 V & $10.38252\pm0.00002$ & $11.0^{+2.1}_{-1.4}$ & $0.89^{+0.18}_{-0.11}{}^{\rm D}$ & $0.079\pm0.007$ & $43.2^{+2.1}_{-2.7}$ & $(\sim4400)^{\rm M}$ & $4.95^{+0.69}_{-0.65}$ & W16, W15 \\
10 & Swift J1727.8-1613\textsuperscript{2} & -- & K4 V & $10.8038\pm0.0010$ & $>3.12$ & -- & -- & $<74$ & $4570^{+90}_{-100}{}^{\rm S}$ & $3.4\pm0.3$ & M25 \\
11 & H 1705-250\textsuperscript{1} & V2107 Oph & K3--M0 V & $12.51\pm0.03$ & $4.9$--$7.9$ & $(0.30$--$0.60)^{\rm C}$ & $\leq0.053$ & $48$--$80$ & -- & $8.6\pm2.1$ & F97, H97, R96 \\
12 & MAXI J1820+070\textsuperscript{1} & V3721 Oph & K3--K5 & $16.4518\pm0.0002$ & $5.71$--$8.10$ & $0.49^{+0.11}_{-0.10}{}^{\rm D}$ & $0.07\pm0.01$ & $67$--$81$ & $(4200$--$4500)^{\rm M}$ & $2.96\pm0.33$ & T19, T20, M22 \\
13 & 4U 1543-475\textsuperscript{1}$^\dagger$ & IL Lup & A2 V & $26.79377\pm0.00007$ & $8.4$--$10.4$ & $(2.30$--$2.60)^{\rm C}$ & $0.25$--$0.31$ & $20.7\pm1.5$ & $9000\pm500^{\rm S}$ & $7.5\pm0.5$ & O98, O03 \\
14 & XTE J1550-564\textsuperscript{1} & V381 Nor & K2--K4 IV & $37.0088\pm0.0001$ & $9.10\pm0.61$ & $0.30\pm0.07^{\rm D}$ & $0.033\pm0.008$ & $74.7\pm3.8$ & $4450^{+200}_{-150}{}^{\rm S}$ & $4.5\pm0.5$ & O11 \\
15 & GX 339-4\textsuperscript{2} & V821 Ara & K1--K2 IV? & $42.21\pm0.01$ & $2.3$--$9.5$ & $(0.50$--$1.40)^{\rm C}$ & $0.18\pm0.05$ & $37$--$78$ & $(4500$--$5000)^{\rm S}$ & $>5$ & H17 \\
16 & GS 1354-64\textsuperscript{2} & BW Cir & G0--G5 III & $61.068\pm0.002$ & $\geq7.6\pm0.7$ & $\geq1.02^{\rm L}$ & $0.12^{+0.03}_{-0.04}$ & $\leq79$ & $(5091$--$5743)^{\rm S}$ & $\sim25$ & C09, C04 \\
17 & GRO J1655-40\textsuperscript{1} & V1033 Sco & F6 IV & $62.926272\pm0.000096$ & $6.0\pm0.4$ & $2.40\pm0.40^{\rm D}$ & $0.42\pm0.03$ & $69\pm2$ & $6150\pm350^{\rm M}$ & $3.2\pm0.2$ & S03, B02, S99 \\
18 & SAX J1819.3-2525\textsuperscript{1}$^\dagger$ & V4641 Sgr & B9 III & $67.6152\pm0.0002$ & $6.4\pm0.6$ & $2.90\pm0.40^{\rm D}$ & $0.455^{+0.045}_{-0.038}$ & $72\pm4$ & $10250\pm300^{\rm S}$ & $6.2\pm0.7$ & M14, O01 \\
19 & GS 2023+338\textsuperscript{1} & V404 Cyg & K3 III & $155.30803\pm0.00005$ & $9.0^{-0.6}_{+0.2}$ & $0.70^{+0.30}_{-0.20}{}^{\rm D}$ & $0.067\pm0.005$ & $67^{+3}_{-1}$ & $(4200$--$4600)^{\rm S}$ & $2.39\pm0.14$ & K10, C94 \\
20 & GRS 1915+105\textsuperscript{1} & V1487 Aql & K1--K5 III & $812\pm4$ & $11.2\pm2.0$ & $0.47\pm0.27^{\rm D}$ & $0.042\pm0.024$ & $60\pm5$ & -- & $9.4\pm0.8$ & R23, S13, R14 \\
\hline
\multicolumn{12}{l}{\textit{B. Indirect and quasi-dynamical constraints}} \\
\hline
C1 & Swift J1357.2-0933\textsuperscript{3} & -- & M2--M4 V & $2.8\pm0.3$ & $12.4\pm3.6$ & -- & $\sim0.04$ & $\sim90$ & -- & $\geq6$ & M15, C13 \\
C2 & MAXI J1659-152\textsuperscript{3} & -- & M2--M5 V & $2.414\pm0.005$ & $3.3$--$7.5$ & -- & $0.02$--$0.07$ & $70$--$80$ & -- & -- & T21 \\
C3 & GRS 1716-249\textsuperscript{3} & Nova Oph 1993 & K6--K7 V & $6.672\pm0.192$ & $6.4^{+3.2}_{-2.0}$ & -- & $0.02$--$0.25$ & $61\pm15$ & -- & $6.9\pm1.1$ & C23 \\
C4 & KY TrA\textsuperscript{3} & A 1524-61 & K2--K5 V & $6.24\pm0.24$ & $5.8^{+3.0}_{-2.4}$ & -- & $\leq0.31$ & $57\pm13$ & -- & $8.0\pm0.9$ & Y24 \\
C5 & Swift J1753.5-0127\textsuperscript{3} & -- & M4--M5 V & $3.26\pm0.02$ & $8.8\pm1.3$ & -- & $0.023\pm0.006$ & $79\pm5$ & -- & $3.9\pm0.7$ & Y25 \\
C6 & Swift J1910.2-0546\textsuperscript{3} & MAXI J1910-057 & M3--M3.5 & $\sim4.5$; $\sim2.25$  & $8$--$11$ & -- & $0.032\pm0.010$ & $13$--$18$ & -- & $2.8$--$4.0$ & C25 \\
\hline
\end{tabular}
}
\vspace{1mm}
\begin{minipage}{0.99\textwidth}
\footnotesize
\noindent
\textit{Notes.}
$q=M_{\rm donor}/M_{\rm BH}$.
The $M_2$ column does not use products of independent extrema of $q$ and
$M_{\rm BH}$. Superscript $\rm D$ denotes a component-mass posterior or
confidence interval published by the dynamical study; $\rm C$ denotes a
source-published stellar or model constraint used in the system analysis; and
$\rm L$ denotes a one-sided dynamical limit.
For $T_{\rm eff}$, superscript $\rm S$ denotes a numerical spectroscopic or
atmosphere-fit constraint and $\rm M$ a value or interval adopted in the
source light-curve, SED, or stellar model. We do not assign a tabulated
temperature when the source gives only a spectral class. Thus the listed
temperatures are source-level constraints on the donor and are not 
calculated here from a uniform spectral-type calibration.
For GRO~J0422+32, the M4--M5~V spectral type
is taken from W00, whereas the approximately 3200~K temperature
was fixed in the light-curve model of C24, which assumed an M2~V
donor. These entries therefore come from different analyses.
System identifications and catalogue-level quantities were cross-checked
against BlackCAT (\textbf{C16}); adopted values are drawn from the
system-specific references listed below. The superscript is an evidence class
for the dynamical inference, not a formal precision rank for every quantity in
a row.
\textsuperscript{1} A donor absorption-line radial-velocity measurement and a
dynamical solution are available, although individual quantities can remain
model dependent.
\textsuperscript{2} Direct or partial dynamical evidence exists but at least
one of $q$, $i$, or $M_{\rm BH}$ is constrained mainly by a broad interval or
a one-sided limit.
\textsuperscript{3} No donor absorption-line $K_2$ measurement is available:
the component constraints depend on disc-emission-line, superhump, dipping,
colour or geometric calibrations. Class-3 systems may be shown with distinct
symbols as evolutionary targets, but are excluded from fits to the direct
dynamical sample.
$^\dagger$ Intermediate- or high-mass donor comparator, retained to show the
broader evolutionary parameter space but excluded from a homogeneous
low-mass donor subsample.
For MAXI J1820+070, the tabulated inclination is the binary-inclination range;
the $63^\circ \pm 3^\circ$ radio-jet orientation is not substituted for it.
For Swift J1910.2-0546, a period of approximately $4.5$ h is favoured,
but an alternative of approximately $2.25$ h and the possibility
of a neutron-star accretor are not yet excluded. The factor-of-two ambiguity may reflect whether the photometric modulation is interpreted as single-wave heating or reflection, or as double-wave
ellipsoidal variability.
\end{minipage}
\vspace{1mm}
\noindent\begin{minipage}[t]{0.99\textwidth}
\footnotesize
\textit{References.}
\textbf{T04} \citealp{Torres2004}; \textbf{K13} \citealp{Khargharia2013}; \textbf{C24} \citealp{Cherepashchuk2024}; \textbf{W00} \citealp{Webb2000}; \textbf{Y22} \citealp{YanesRizo2022}; \textbf{F99} \citealp{Filippenko1999}; \textbf{S96} \citealp{Shahbaz1996}; \textbf{O04} \citealp{Orosz2004}; \textbf{V17} \citealp{VanGrunsven2017}; \textbf{C10} \citealp{Cantrell2010}; \textbf{C91} \citealp{Charles1991}; \textbf{C95} \citealp{Casares1995}; \textbf{I04} \citealp{Ioannou2004}; \textbf{H96} \citealp{Harlaftis1996}; \textbf{M21} \citealp{MataSanchez2021}; \textbf{W16} \citealp{Wu2016}; \textbf{W15} \citealp{Wu2015}; \textbf{M25} \citealp{MataSanchez2025}; \textbf{F97} \citealp{Filippenko1997}; \textbf{H97} \citealp{Harlaftis1997}; \textbf{R96} \citealp{Remillard1996}; \textbf{T19} \citealp{Torres2019}; \textbf{T20} \citealp{Torres2020}; \textbf{M22} \citealp{Mikolajewska2022}; \textbf{O98} \citealp{Orosz1998}; \textbf{O03} \citealp{Orosz2003}; \textbf{O11} \citealp{Orosz2011}; \textbf{H17} \citealp{Heida2017}; \textbf{C09} \citealp{Casares2009}; \textbf{C04} \citealp{Casares2004}; \textbf{S03} \citealp{Shahbaz2003}; \textbf{B02} \citealp{Beer2002}; \textbf{S99} \citealp{Shahbaz1999}; \textbf{M14} \citealp{MacDonald2014}; \textbf{O01} \citealp{Orosz2001}; \textbf{K10} \citealp{Khargharia2010}; \textbf{C94} \citealp{Casares1994}; \textbf{R23} \citealp{Reid2023}; \textbf{S13} \citealp{Steeghs2013}; \textbf{R14} \citealp{Reid2014}; \textbf{M15} \citealp{MataSanchez2015}; \textbf{C13} \citealp{CorralSantana2013}; \textbf{T21} \citealp{Torres2021}; \textbf{C23} \citealp{Casares2023}; \textbf{Y24} \citealp{YanesRizo2024}; \textbf{Y25} \citealp{YanesRizo2025}; \textbf{C25} \citealp{CorralSantana2025}; \textbf{C16} \citealp{CorralSantana2016}.
\end{minipage}
\end{table*}

\section{Models: initial binary populations and detailed evolution}
\label{sec:models}
Throughout this work, ``donor''
denotes the Roche-lobe filling, mass-losing non-compact star,
while the black hole is the accretor.

\subsection{Initial binary populations}
\label{subsec:initial_populations}

\begin{figure*}
\centering
\includegraphics[width=\linewidth]{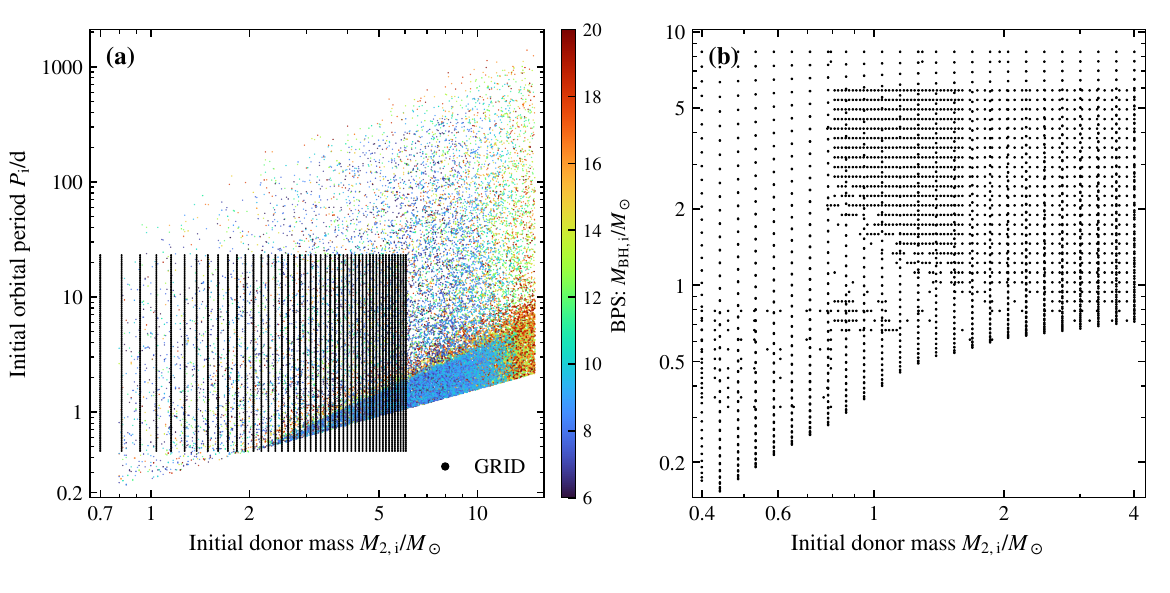}
\caption{Initial donor-mass and orbital-period
coverage. Panel (a) combines the 40,487 regular GRID and 38,308 BPS-seeded irradiation-free tracks. Colour indicates the
initial black-hole mass of the BPS models. GRID models with
different black-hole masses share donor-mass--period coordinates;
these coincident positions are shown once in black to keep the
regular sampling visible against the BPS distribution.
Panel (b) shows the 4,392 irradiated tracks, comprising
2,181 with an initial black-hole mass of \(5\,M_\odot\) and
2,211 with \(8\,M_\odot\). Exactly coincident donor-mass--period
positions in this panel are also represented by a single marker.
The panels have different axis limits, with orbital periods in days
in both panels. Point density reflects the adopted sampling and
is not a formation probability or population weight.}
\label{fig:Fig01}
\end{figure*}

The two irradiation-free initial populations fall into two categories.
The first is a regular grid in black-hole mass, donor mass and orbital
period (hereafter GRID). The second is generated by a simplified binary
population-synthesis calculation (BPS). 
Fig.~\ref{fig:Fig01}(a) shows the initial GRID
and BPS samples together in the
\(P_{\rm orb}\)--\(M_2\) plane. 
The GRID set is not assigned a Galactic
population weight.

The BPS calculation adopts
\(\alpha_{\rm CE}\lambda=0.20\), and neglects wind-driven orbital
expansion before the common-envelope phase. Primordial primary masses are drawn from a Salpeter initial mass
function, $dN/dM_1\propto M_1^{-2.35}$, truncated to
$18\leq M_1/M_\odot\leq45$ \citep{Salpeter1955}. The initial mass
ratio is sampled uniformly before the retained companion-mass cuts
are applied. The code assigns a core mass to the primary at common-envelope entry
using $M_{\rm core}=0.12M_1^{1.35}$, a scaling given by
\citet{Podsiadlowski2003}. Here $M_1$ is the zero-age main-sequence
primary mass and all masses are in solar units. The simplified BPS
calculation then identifies this core mass with the final black-hole
mass, $M_{\rm BH}=M_{\rm core}$. 
The simplified BPS code further approximates the initially unevolved
companion radius by \(R_2=0.8 M_2^{0.85}\), where the quantities are
again in solar units. Comparisons with the results of the HPT/BSE code
\citep{Hurley2000,Hurley2002} show that neglecting orbital expansion
can allow CE entry in systems that would otherwise avoid it, while
the smaller assumed companion radii in the examined cases near
\(1.4\,M_\odot\) can favour survival in post-CE contact tests.
For the same primordial inputs and CE parameters, the BPS sample has
a more pronounced low-donor-mass tail and extends further towards
long post-CE periods than a comparable HPT/BSE sample,
although their median periods are similar. 

We use this BPS set as an illustrative, nonuniform extension of the GRID sample to explore the subsequent evolution, conditional on the adopted BH mass, donor mass, and orbital period distributions in the incipient population. The two sets (GRID vs BPS) are distinguished throughout, and the BPS
sampling density is not interpreted as a formation probability or as
evidence that low-mass companions physically survive CE evolution.

\subsection{Parameter space and physical motivation}
The parameter ranges are chosen to encompass both classical low-mass X-ray binaries and systems potentially arising from intermediate-mass X-ray binary (IMXB) progenitors. For the regular GRID sample, black-hole masses span $4\leq M_{\rm BH}/M_\odot \leq 20$, while donor masses extend from $M_{\rm donor}\simeq0.7\,M_\odot$ to $6\,M_\odot$, thereby covering the regime most relevant to BH--LMXB and stripped-IMXB evolution \citep[e.g.][]{Podsiadlowski2003,Li2015}. Initial orbital periods in the GRID sample cover a broad range, from fractions of a day to tens of days, capturing both compact post-common-envelope systems and wider binaries that may shrink by angular-momentum loss. The BPS-seeded sample extends to a broader range of initial donor masses and orbital periods, including a tail toward higher donor masses, but this broader population is used primarily to test the sensitivity of the results to the initial binary distribution rather than to redefine the main BH--LMXB parameter domain considered here. Although the BPS-seeded sample extends to higher donor masses, the present analysis focuses on systems evolving through the BH--LMXBs and stripped-donor channels. Possible evolutionary pathways leading to BH--NS binaries are not explicitly followed here and will be explored in future work.

In many systems, particularly those with donor masses greater than about $1.2\,M_\odot$, Roche-lobe overflow is initiated primarily by the nuclear expansion of the donor star rather than by orbital shrinkage. As shown in Fig.~\ref{fig:Fig01}, this broad sampling avoids imposing preconceptions on which regions of parameter space are allowed to evolve into BH--LMXBs. Instead, it enables us to determine, in a controlled manner, which initial configurations actually populate the observed planes once detailed binary evolution is followed, with the onset of Roche-lobe contact and subsequent orbital evolution governed by the angular-momentum-loss and mass-transfer response formalism introduced below.

This broadly sampled parameter space addresses persistent challenges in BH--LMXB formation. The survival of low-mass companions through the common-envelope phase remains energetically problematic \citep{Kalogera1999}, whereas IMXB progenitors more readily survive but can yield discrepancies in donor effective temperatures at later stages \citep{Justham2006,Fan2024}. By including both regimes, the grid allows us to examine how these channels are represented in the resulting distributions, even though a detailed mapping between specific initial configurations and final evolutionary tracks is not pursued in this work.

\subsection{Model Set, Parameter Coverage, and Sampling}
\label{sec:model_sampling}
To quantify the scope and resolution of our binary evolution calculations, we summarize here the size, parameter coverage, and sampling properties of the model sets introduced above.

\subsubsection{Irradiation-free reference populations}
\label{subsubsec:noirrad_sampling}
After the initial file conversion but before the subsequent
inspection for duplicated evolutionary histories, the analysis tables
contained \(79{,}675\) track identifiers, \(40{,}487\) GRID and
\(39{,}188\) BPS tracks. These counts were formed after conversion of
the original model outputs, during which no GRID files and \(389\) BPS
files had already been excluded as malformed or unusable. A subsequent
audit identified \(880\) additional BPS entries as exact four-row copies
of an unrelated sequence. These entries are excluded from every analysis
of physical evolutionary histories. 
The pre-audit converted archive is retained unchanged for provenance,
whereas all downstream analysis products and reported physical counts
are regenerated after applying this duplicate mask.
The usable irradiation-free sample
therefore contains \(78{,}795\) tracks, \(40{,}487\) GRID and
\(38{,}308\) BPS tracks. The corresponding usable history tables contain
\(33{,}939{,}541\) GRID and \(33{,}066{,}764\) BPS rows. Ten of the
retained BPS tracks contain valid late-evolution suffixes but not their
complete leading histories. Their stored states are retained in
state-space maps but they are omitted from statistics that require
complete histories.

The GRID models provide a controlled and uniformly sampled exploration
of parameter space, while the BPS-seeded models provide a complementary,
dense, and more irregular sampling of
post-formation black-hole--donor initial conditions.

For the GRID suite, the initial parameter domain is
\begin{center}
\resizebox{0.98\columnwidth}{!}{%
  $\displaystyle
  M_{2,\mathrm{init}} \in [0.7,\,6.013]\,M_\odot,\quad
  P_{\mathrm{init}} \in [0.468,\,22.746]\,\mathrm{d},\quad
  M_{\mathrm{BH,init}} \in [4,\,20]\,M_\odot
  $.%
}
\end{center}
The GRID sampling is approximately regular in donor mass and BH mass,
\[
\Delta M_{2}\simeq 0.113\,M_\odot,\qquad
\Delta M_{\mathrm{BH}}=2\,M_\odot,
\]
with \(48\) donor masses and \(9\) black-hole masses. The initial periods comprise \(100\) approximately logarithmically spaced values, with
\[
\Delta\log_{10}P \approx 0.017~\mathrm{dex}.
\]
For the BPS suite, the initial parameter domain is
\[
\begin{gathered}
M_{2,\mathrm{init}} \in [0.8,\,15.0]\,M_\odot,\qquad
P_{\mathrm{init}} \in [0.237,\,1377]\,\mathrm{d},\\
M_{\mathrm{BH,init}} \in [6,\,20]\,M_\odot.
\end{gathered}
\]
These ranges describe the post-population-synthesis
BH--donor binaries used as initial conditions for the subsequent
\textsc{MESA} calculations; they are not the primordial binary
parameters. The BPS-seeded sample is treated as an unweighted,
irregular sampling of the post-formation parameter space.

Beyond the file-integrity exclusions described above, no
selection on physical outcome is applied. All successfully computed
usable models are retained.

These two irradiation-free samples form the basis of
Figs~\ref{fig:Fig02}--\ref{fig:Fig07} and of the gravitational-wave
analysis in Section~\ref{sec:compact_lisa}.

\subsubsection{Irradiated sample}
\label{subsubsec:irrad_sampling}

This sample explores how irradiation affects black-hole
binary evolution. It contains 4,392 usable irradiated tracks:
2,181 with an initial black-hole mass of $5\,M_\odot$ and
2,211 with $8\,M_\odot$. 
Three illustrative histories selected from this sample are examined in Section~\ref{sec:detailed_irradiation}.

The irradiated sample was calculated on
TRUBA\footnote{TRUBA is the Turkish National e-Science
e-Infrastructure, operated by T\"{U}B\.{I}TAK ULAKB\.{I}M.},
Fir\footnote{Fir is a Digital Research Alliance of Canada
supercomputer hosted at Simon Fraser University.}, and
CSD3\footnote{Cambridge Service for Data Driven Discovery,
operated by the University of Cambridge Research Computing
Service.}. All calculations in this sample use the same
irradiation prescription.

The following partly overlapping initial-condition ranges
are combined to improve coverage near the bifurcation region and
the onset of Roche-lobe overflow, including both initially
low-mass donors and IMXB progenitors.
The initial conditions span three partly overlapping
donor-mass--orbital-period ranges:
\[
\begin{aligned}
M_{2,\mathrm{init}}&\in[0.40,4.00]\,M_\odot,
& P_{\mathrm{init}}&\in[3.71,200]\,\mathrm{h},\\
M_{2,\mathrm{init}}&\in[0.81,4.00]\,M_\odot,
& P_{\mathrm{init}}&\in[14.71,141.22]\,\mathrm{h},\\
M_{2,\mathrm{init}}&\in[0.78,1.95]\,M_\odot,
& P_{\mathrm{init}}&\in[19.09,200]\,\mathrm{h}.
\end{aligned}
\]
\par

For the first range, a mass-dependent lower-period
boundary, $P_{\rm min}=1.05P_{\rm RL95}$, was imposed to avoid
initial Roche-lobe overflow. Here $P_{\rm RL95}$ is the
interpolated period at which the donor fills 95 per cent of
its Roche lobe. Periods extend to $200\,\mathrm{h}$, with
enhanced sampling near $P_{\rm min}$. The second selection
was drawn from an adaptive parent grid after excluding
targets expected to start with Roche-lobe overflow and
exact or near duplicates of previous calculations.
The third range contributes additional
distinct initial configurations not already represented by
the first two selections.
Repeated calculations of the same target on different
facilities are counted only once in the combined sample.

The three mass--period ranges partly overlap near
the bifurcation region and the onset of Roche-lobe overflow.
Here the bifurcation region denotes the neighbourhood of the
mass-dependent initial orbital period that separates
predominantly converging systems, in which angular-momentum
loss drives the binary towards shorter periods, from diverging
systems, in which the donor's nuclear expansion drives evolution
towards longer periods
\citep{PylyserSavonije1988,PylyserSavonije1989}.
The location of this boundary is sensitive to the donor's
evolutionary state and to the adopted angular-momentum-loss
prescription \citep{NelsonDubeau2004,GoliaschNelson2015}.
It does not represent a boundary between LMXBs and IMXBs;
the sampled region includes both initially low-mass donors
and IMXB progenitors.

The relative numbers of calculations
in these ranges reflect the sampling design and are not used
as formation-rate weights.
Dedicated follow-up calculations are kept outside this sample. 
The combined initial coverage of the irradiated
sample is shown in Fig.~\ref{fig:Fig01}(b).

\subsection{Irradiation and the state-dependent low-efficiency prescription}
\label{subsec:irradiation_method}

The irradiation-free GRID and BPS calculations described
above do not include the effects of irradiation of the donor star by
X-rays and UV radiation produced by the black-hole accretion disc. In
the targeted irradiated production grids, this irradiation is explicitly
included and directly affects the donor's outer layers. The accretion
luminosity is evaluated as
\begin{equation}
L_{\rm X}=\eta_{\rm acc}\dot M_{\rm acc}c^2,
\qquad \eta_{\rm acc}\simeq0.1,
\end{equation}
where $\dot{M}_{\rm acc}$ is limited by the Eddington
accretion rate. The intercepted heating luminosity is
\begin{equation}
L_{\rm heat}
 =\eta_{\rm irr}\frac{R_2^2}{4a^2}L_{\rm X},
\label{eq:irradiation_heating}
\end{equation}
where \(R_2\) is the donor radius, \(a\) is the orbital
separation, and \(\eta_{\rm irr}\) is the fraction of the geometrically
intercepted accretion luminosity that is deposited in the donor's outer
layers. These calculations therefore include
feedback between accretion luminosity, donor heating, stellar radius, and
Roche-lobe overflow.

The intercepted luminosity is deposited through the donor's outer
optical-depth layers using the custom \texttt{run\_star\_extras}
implementation. Between adjacent mesh points, the remaining heating
luminosity is attenuated by
\(\exp[-(\tau_k+\tau_{k+1})/2]\), and the absorbed fraction is added as
specific heating divided by the shell mass. Here \(k\) labels adjacent
stellar mesh points and \(\tau_k\) is the optical depth at mesh point
\(k\). Thus the irradiation prescription modifies the donor star's
energy equation rather than being applied only as a post-processing
luminosity correction.
The absorbed energy is therefore not forced to be reradiated
instantaneously: it can alter the donor's local thermal structure,
radius, and Roche-lobe overflow rate before emerging from the surface.
By contrast, simply adding the absorbed power to the reported stellar
luminosity would not feed back on the binary evolution. 

We adopt a nominal irradiation efficiency of
\(\eta_0=0.05\). We assume that most of the solid angle subtended by
the donor is shadowed by the accretion disc, so that only about
10 per cent of its visible surface is irradiated, and that approximately
half of the incident radiation penetrates deeply enough to affect the
outward energy flow. The resulting efficiency is necessarily uncertain,
but it is sufficient to produce a measurable effect on the short-term
binary evolution.

At this efficiency, calculations of the stellar structure,
mass-transfer rate, and binary orbit become computationally very
expensive as the donor mass approaches \(0.4\,M_\odot\) on the
short-period branch. Rather than retain the nominal irradiation
efficiency while following these systems to very low donor masses and
ultrashort orbital periods, we reduce the irradiation efficiency
according to the following criteria. In successful paired pilot
continuations, \(\eta_{\rm low}=0\) and \(10^{-5}\) yielded nearly
indistinguishable tracks in the \(P_{\rm orb}\)--\(M_2\) plane. We
therefore adopt \(\eta_{\rm low}=10^{-5}\) as a near-zero numerical
floor: irradiation is then effectively negligible, while the
prescription remains formally non-zero and does not rely on an
exact-zero special case. The code evaluates the effective irradiation
efficiency at every model,

\begin{equation}
\eta_{\rm irr,eff}=
\begin{cases}
\eta_{\rm low}=10^{-5}, & \mathrm{if}\,\, \mathcal{C}_{\rm low},\\
\eta_0=0.05, & \mathrm{otherwise},
\end{cases}
\label{eq:irradiation_efficiency}
\end{equation}
where
\begin{align}
\mathcal{C}_{\rm low}={}&
\left(M_2<0.4\,M_\odot \ \land\ P_{\rm orb}<4~\mathrm{h}\right)
\nonumber\\
&{}\lor
\left(M_2<0.45\,M_\odot \ \land\ P_{\rm orb}<4~\mathrm{h}
\ \land\ \Delta t<2\times10^{-9}~\mathrm{yr}\right).
\label{eq:low_efficiency_condition}
\end{align}
The condition is evaluated at every model. A saved
logical flag records only the first entry into the low-efficiency regime for
logging purposes; it does not permanently latch the subsequent evolution at
$\eta_{\rm low}$. We therefore refer to the affected portions of a track as
state-dependent low-efficiency segments rather than as a post-switch
continuation. The term ``irradiated production grid'' denotes the calculation
campaign as a whole and does not imply that every model along every track was
evolved with a constant $\eta_{\rm irr}=0.05$. Minimum-period and compact-branch
results are identified explicitly when they occur in the low-efficiency
regime.

The targeted irradiated calculations stop at the adopted
maximum age or low-mass limit, or when the timestep falls below the
adopted numerical floor. At very low donor masses the calculations
become increasingly expensive, and some tracks end before the complete
short-period tail is resolved. Their converged portions are retained,
but a numerical stopping point is not interpreted as a physical
endpoint.

\subsection{Detailed binary evolution with \textsc{MESA}}
\label{subsec:relations}

The 78,795-track irradiation-free GRID and
BPS-seeded reference populations were evolved with \textsc{MESA}
revision 3851, the same revision used for the cataclysmic-variable
calculations presented by \citet{Kalomeni2016}.
The targeted irradiated production grids were evolved with
\textsc{MESA} release 26.4.1.
Three selected irradiated models were
recalculated with the same release, initial conditions,
mass-retention prescription and other input controls, but with
the irradiation efficiency fixed at
\(\eta_{\rm irr}=10^{-10}\).
These three negligible-irradiation calculations provide the
comparison curves in Fig.~\ref{fig:irradiation}; they are not
included in the irradiation-free reference samples or the
irradiated sample counts.

All detailed binary calculations adopt initially circular
orbits, \(e_{\rm init}=0\), representing tidally circularized
post-formation systems. Where magnetic braking is applied, the donor
rotation is assumed to be synchronized with the orbit.

Mass transfer via Roche-lobe overflow is modelled directly. In the
irradiation-free GRID and BPS calculations, the legacy custom implementation
sets the retained fraction of transferred material to unity. The targeted
irradiated production grids instead use the non-conservative prescription
specified below. The Roche-lobe radius $R_{\rm L}$ is approximated, following \citet{Eggleton1983}, as
\begin{equation}
\frac{R_{\rm L}}{a} = \frac{0.49\,q^{2/3}}{0.6\,q^{2/3} + \ln(1+q^{1/3})},
\label{eq:eggletonRL}
\end{equation}
where $a$ is the orbital separation and $q = M_{\rm donor}/M_{\rm BH}$ is the mass ratio.

Roche-lobe overflow occurs when $R_{\rm donor}\simeq R_{\rm L}$.
Together with Kepler's third law, this relation provides a direct
physical interpretation of the loci traced in the
$P_{\rm orb}$--$M_{\rm donor}$ and $q$--$P_{\rm orb}$ planes.

In the targeted irradiated production calculations, the donor is evolved
as a stellar model and the black hole is represented as a point-mass
accretor. We calculate the Roche-lobe overflow rate using the optically
thick mass-transfer prescription of \citet{Kolb1990}. In the MESA
mass-loss convention, \(\alpha_{\rm MT}\) denotes the fraction of the
transferred mass lost from the vicinity of the donor as a fast wind,
\(\beta_{\rm MT}\) denotes the fraction lost from the vicinity of the
accretor, and \(\delta_{\rm MT}\) denotes
the fraction lost through a circumbinary coplanar toroid. We adopt
\(\alpha_{\rm MT}=0\), \(\beta_{\rm MT}=0.7\), and
\(\delta_{\rm MT}=0\). The value \(\beta_{\rm MT}=0.7\) was adopted to
reproduce the benchmark setup used for the targeted irradiated
calculations. In our MESA convention, this corresponds to the fixed
nominal retained fraction
\[
f_0=1-\alpha_{\rm MT}-\beta_{\rm MT}-\delta_{\rm MT}=0.3.
\]
An effective accretion efficiency of 30 per cent has previously been
adopted in neutron-star LMXB calculations
\citep{TaurisLangerKramer2012}, but we do not regard that value as a
calibration for BH--LMXBs. It is therefore treated here as a modelling
assumption. MESA can reduce the actual black-hole mass growth further
through the Eddington limit and radiative-efficiency correction.

Angular-momentum losses include gravitational radiation,
mass-loss terms, and the standard MESA implementation of the Rappaport,
Verbunt \& Joss magnetic-braking prescription with
$\gamma_{\rm MB}=3$. 
Magnetic braking is active when the donor has a surface-connected
convective envelope with mass fraction
\(q_{\rm conv,env}\equiv M_{\rm conv,env}/M_{\rm donor}\) satisfying
\(10^{-6}<q_{\rm conv,env}<0.99\), and no significant convective core,
where \(q_{\rm conv,core}\equiv M_{\rm conv,core}/M_{\rm donor}<10^{-2}\).
The threshold is imposed on the instantaneous
convective-envelope mass fraction rather than on a fixed donor mass;
the corresponding donor mass therefore depends on its evolutionary
state. For $q_{\rm conv,env}<0.02$, the standard low-envelope scaling is applied.
The MESA control \texttt{keep\_mb\_on} determines whether magnetic
braking is forced to remain active after the donor loses its radiative
core. We set \texttt{keep\_mb\_on=.false.}, so the default
structure-dependent braking criterion is applied and magnetic braking
is not forced to continue once the donor becomes fully convective.
The total fractional angular-momentum loss rate is expressed as
\begin{equation}
\frac{\dot{J}}{J} = \left(\frac{\dot{J}}{J}\right)_{\rm GR} 
+ \left(\frac{\dot{J}}{J}\right)_{\rm MB} 
+ \left(\frac{\dot{J}}{J}\right)_{\rm ML}
\label{eq:Jdot_total}
\end{equation}
where the gravitational radiation term follows \citet{Landau1975}, and magnetic braking is implemented according to \citet{Rappaport1983}. Magnetic braking is applied only to donors possessing convective envelopes and a radiative core. The additional term $(\dot{J}/J)_{\rm ML}$ accounts for angular-momentum loss associated with non-conservative mass transfer. Material ejected through \(\beta_{\rm MT}\), or by the Eddington
limit, is assumed to leave from the vicinity of the black hole,
carrying the black hole's specific orbital angular momentum
\citep{TaurisSavonije1999}.

Gravitational radiation operates throughout the calculations
and provides a baseline contribution to angular-momentum loss at short
orbital periods. Magnetic braking supplies an additional torque while
the donor satisfies the adopted structural criteria and may dominate
the orbital evolution during those phases. Once magnetic braking
switches off, gravitational radiation remains active.

The orbital response to mass transfer can be written as
\citep{Podsiadlowski2003}
\begin{equation}
\frac{\dot{a}}{a} =
-2\frac{\dot{M}_{\rm d}}{M_{\rm d}}
\left[
1-f_{\rm acc}q
-\frac{1-f_{\rm acc}}{2}\frac{q}{1+q}
\right]
+2\frac{\dot{J}}{J},
\label{eq:orbital_response}
\end{equation}
where $\dot{M}_{\rm d}<0$ is the donor mass-loss rate and
\[
f_{\rm acc}\equiv-\frac{\dot{M}_{\rm BH}}{\dot{M}_{\rm d}}
\]
is the instantaneous fraction of the transferred material retained by the
black hole. This time-dependent retained fraction is distinct from the fixed
MESA mass-loss parameter \(\beta_{\rm MT}\). 

To interpret the secular evolution, it is useful to write the orbital-period derivative in the general form

\begin{equation}
\frac{\dot{P}_{\rm orb}}{P_{\rm orb}} = 3\frac{\dot{J}}{J}
-3\left(\frac{\dot{M}_{\rm BH}}{M_{\rm BH}}+\frac{\dot{M}_{\rm donor}}{M_{\rm donor}}\right)
+\frac{\dot{M}_{\rm bin}}{M_{\rm bin}},
\label{eq:Porb_general}
\end{equation}

\noindent where $M_{\rm bin}=M_{\rm BH}+M_{\rm donor}$ is the total binary mass,
$J$ is the orbital angular momentum, and overdots denote time derivatives.
In our calculations, $\dot{J}$ includes gravitational radiation,
magnetic braking where applicable, and the orbital angular momentum
carried away by escaping matter. Mass transfer and systemic mass
loss also determine the component-mass and total-mass derivatives
in equation~\eqref{eq:Porb_general}.

At short orbital periods, gravitational radiation provides a robust and model-independent baseline contribution to angular-momentum loss.  
Within the quadrupole approximation, this contribution can be written as \citep{Landau1975}

\begin{equation}
\left(\frac{\dot{J}}{J}\right)_{\rm GR}
= -\frac{32}{5}\frac{G^3}{c^5}\frac{M_{\rm BH}M_{\rm donor}(M_{\rm BH}+M_{\rm donor})}{a^4}.
\label{eq:Jdot_GR}
\end{equation}

For systems with non-degenerate, magnetically active donor stars,
gravitational radiation alone is generally insufficient to drive the
observed long-term mass-transfer rates. Magnetic braking can arise from
a rotation-driven dynamo and a magnetically coupled stellar wind
\citep{Tout1992}. In a synchronised close binary, tidal coupling
transfers angular momentum from the orbit to the stellar spin, allowing
the wind torque to drain the orbit \citep{Yakut2005}. We do not
evolve this dynamo--wind coupling explicitly. When the donor satisfies
the structural criteria described above, we instead use the classical
prescription of \citet{Rappaport1983},

\begin{equation}
\dot{J}_{\rm MB}
= -3.8\times10^{-30}\, M_{\rm donor}\,R_\odot^{4}
\left(\frac{R_{\rm donor}}{R_\odot}\right)^{\gamma}
\omega^{3}.
\label{eq:Jdot_MB}
\end{equation}

Here $R_{\rm donor}$ is the donor radius and $\omega$ is the
orbital angular velocity, assuming synchronous rotation. The production
calculations use the standard MESA implementation of Rappaport, Verbunt
\& Joss' prescription without an inlist override of the magnetic-braking
exponent. The effective value is therefore
$\gamma=\gamma_{\rm MB}=3$. 

Under the adopted MESA criteria, magnetic braking is inactive
initially for donors more massive than approximately
\(1.3\,M_\odot\), whose outer envelopes are predominantly radiative
and it also switches off when the donor later becomes fully convective
or degenerate. The full-convection switch resembles, but is not
identical to, the two-dynamo picture of \citet{Zangrilli1997}. In their 
model the core--envelope boundary dynamo disappears when the radiative
core vanishes, while a weaker envelope dynamo remains. We do not
implement a double dynamo prescription here.

X-ray irradiation is omitted from the 78,795-track
irradiation-free GRID and BPS reference populations, but is included explicitly
in the separate targeted production grids through the prescription described
in Section~\ref{subsec:irradiation_method}.

For the irradiation-free GRID and BPS reference populations, the adopted
stopping conditions include permanent detachment and a donor-mass floor
of approximately \(0.01\,M_\odot\). Some sequences also terminate at
\(P_{\rm orb}=1000\,\mathrm{h}\) while their stripped donors remain
extended. The last stored models therefore do not necessarily represent
physical endpoints.

\subsection{Scope and limitations of the modelling framework}
Together, the GRID and BPS reference samples show how the secular morphology
depends on the adopted initial sampling, whereas the smaller targeted
grids examine the donor's structural and mass-transfer response to
irradiation. The present \textsc{MESA} calculations follow the secular evolution
of post-formation BH--donor binaries under the adopted mass-transfer,
angular-momentum loss, and irradiation prescriptions; they do not model
black-hole formation or common-envelope ejection.

The calculations do not evolve the thermal--viscous instability of the
accretion disc and therefore do not resolve outburst--quiescence duty
cycles. The accretion rate used in the donor-irradiation prescription
is the rate resolved on the stellar-evolution timestep, not an outburst
accretion rate. In representative secular portions of the calculations,
the \textsc{MESA} timesteps are of order \(10^5\) yr, although the
adaptive timestep can become much shorter during rapid or numerically
difficult phases; these timescales remain distinct from the
disc-instability cycles, of order a decade, that are not resolved here.
We consequently do not classify individual tracks as persistent or
transient. Such a classification would require an additional
irradiated-disc instability prescription, as used, for example, by
\citet{DengLi2024}.

\section{Mapping the Model Grid onto Observational Planes}
\label{sec:mapping}

We compare the evolutionary tracks with the observed systems in the
\(P_{\rm orb}\)--\(M_{\rm donor}\), \(q\)--\(P_{\rm orb}\), and
\(P_{\rm orb}\)--\(T_{\rm eff}\) planes.
For the mapping analysis, each MESA sequence is followed from
the onset of Roche-lobe overflow to its final converged model.
Temporary detached intervals, occurring after the first RLOF episode, are
retained as part of the subsequent binary evolution. For each stored model we
record the observable quantities $(P_{\rm orb},M_{\rm donor},T_{\rm eff})$,
the mass ratio $q=M_{\rm donor}/M_{\rm BH}$ and the physical time interval
to the following stored model. Each binary therefore contributes an
evolutionary locus rather than a single endpoint. The construction and
interpretation of the time-weighted maps are defined below.

In each parameter plane, we overlay the
dynamically confirmed Galactic BH--LMXB sample, using the
observational parameters and uncertainties described in
Section~\ref{sec:observedBHs}. The comparison is carried out
directly in the observable quantities, avoiding derived
evolutionary variables and thereby limiting additional assumptions.

\subsection{Construction of time-weighted evolutionary maps}
\label{subsec:time_weighting}

For each evolutionary track \(k\) and stored
evolutionary step \(i\), let \((x_{ki},y_{ki})\) denote the model's
horizontal and vertical coordinates in the plotted plane
(donor mass and orbital period, mass ratio and orbital period,
or orbital period and effective temperature). The coordinates
include any logarithmic transformations used in the corresponding
map.

We label each two-dimensional bin by its centre \((x,y)\).
Let \(\mathcal{A}_k\) denote the indices of the finite, positive
time intervals between consecutive stored models. The time
contributed by track \(k\) to this bin is
\begin{equation}
\begin{aligned}
\Delta t_{ki} &= t_{k,i+1}-t_{k,i},\\
T_k(x,y) &= \sum_{i\in\mathcal{A}_k}
\Delta t_{ki}\,\mathcal{I}_{x,y}(x_{ki},y_{ki}),
\end{aligned}
\label{eq:track_bin_time}
\end{equation}
where \(\mathcal{I}_{x,y}=1\) when the stored model lies in the
bin centred at \((x,y)\), and is zero otherwise. Summing over the
\(N\) tracks gives the cumulative time in that bin,
\begin{equation}
T_{\rm bin}(x,y)=\sum_{k=1}^{N}T_k(x,y).
\label{eq:raw_bin_time}
\end{equation}
For panels displaying a relative dwell-time fraction, we use
\begin{equation}
F(x,y)=
\frac{T_{\rm bin}(x,y)}
{\displaystyle\sum_{(x',y')\in\mathcal{B}}T_{\rm bin}(x',y')},
\label{eq:normalised_occupancy}
\end{equation}
where \(\mathcal{B}\) is the set of bin centres in the displayed
panel. Thus \(T_{\rm bin}(x,y)\) is the total computed time spent
in that bin, while \(F(x,y)\) is its fraction of the total time
represented in the panel. No binary-formation or
observational-selection weights are applied, so these maps do
not represent an absolute Galactic source density.

\section{Results from the Irradiation-free Reference Populations}
\label{sec:noirrad_results}
\subsection{Orbital-period--donor-mass evolution}
\label{subsec:period_donor_results}

\begin{figure*}
\centering
\includegraphics[width=0.95\linewidth]{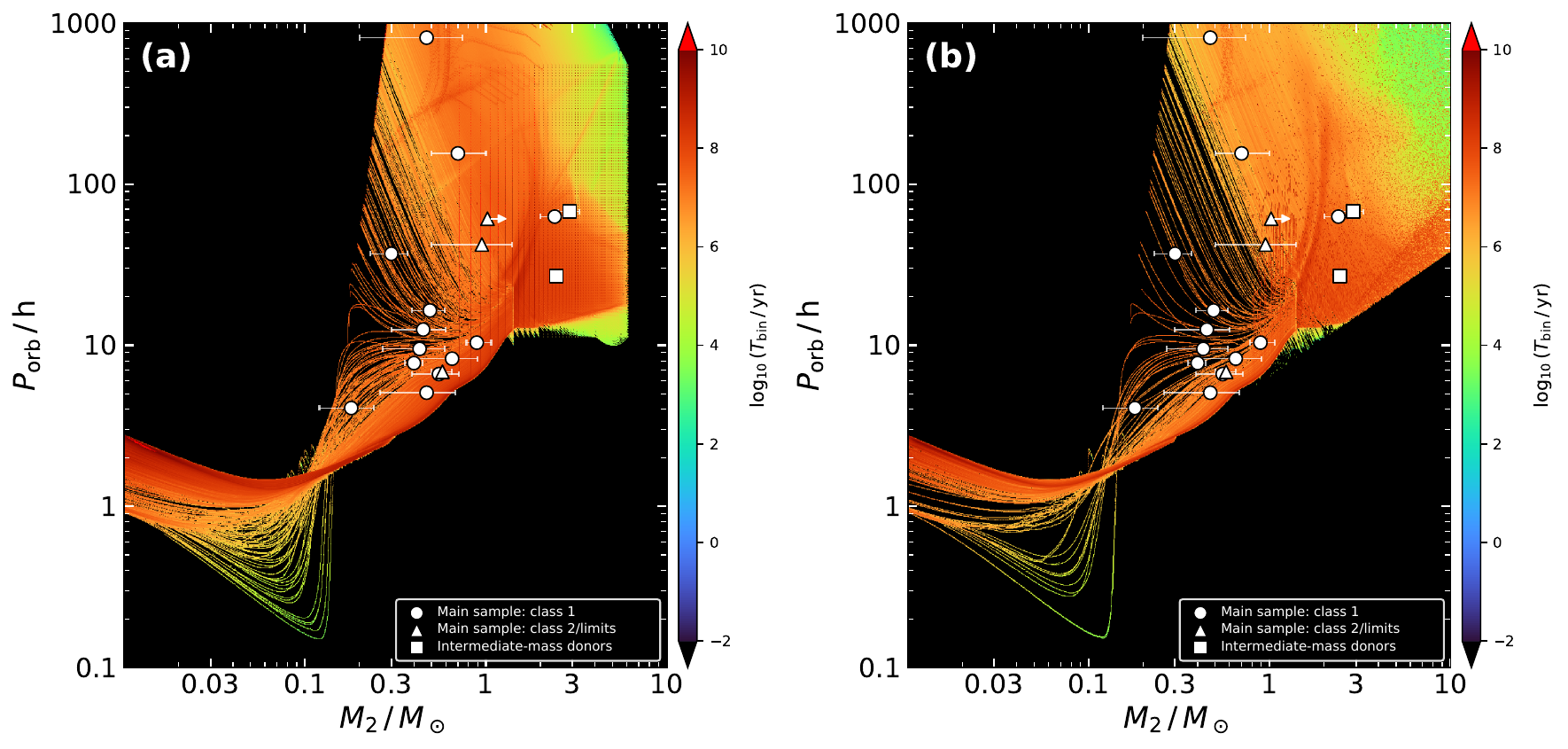}
\caption{Evolutionary tracks projected onto the
orbital-period--donor-mass plane for (a) the GRID and (b) the BPS-seeded
irradiation-free calculations. Each occupied pixel is colour-coded by
\(\log_{10}[T_{\rm bin}/{\rm yr}]\), where \(T_{\rm bin}\) is the raw
cumulative computed evolutionary time defined in
equation~\ref{eq:raw_bin_time}. This quantity is not normalized per track
and therefore depends on the number, duration and initial coverage of
the calculated sequences. It is a numerical dwell-time diagnostic
rather than a formation-weighted population density.
White circles and triangles distinguish class-1 and class-2
systems, respectively, in the main observational sample. White squares
denote the two additional systems with intermediate-mass donors;
this symbol identifies their role as comparison systems rather than
an additional dynamical-evidence class.
Error bars show the published donor-mass uncertainties and arrows
denote one-sided limits. Class-3 systems are not included in this
overlay.}
\label{fig:Fig02}
\end{figure*}

\begin{figure*}
\centering
\includegraphics[width=0.95\linewidth]
{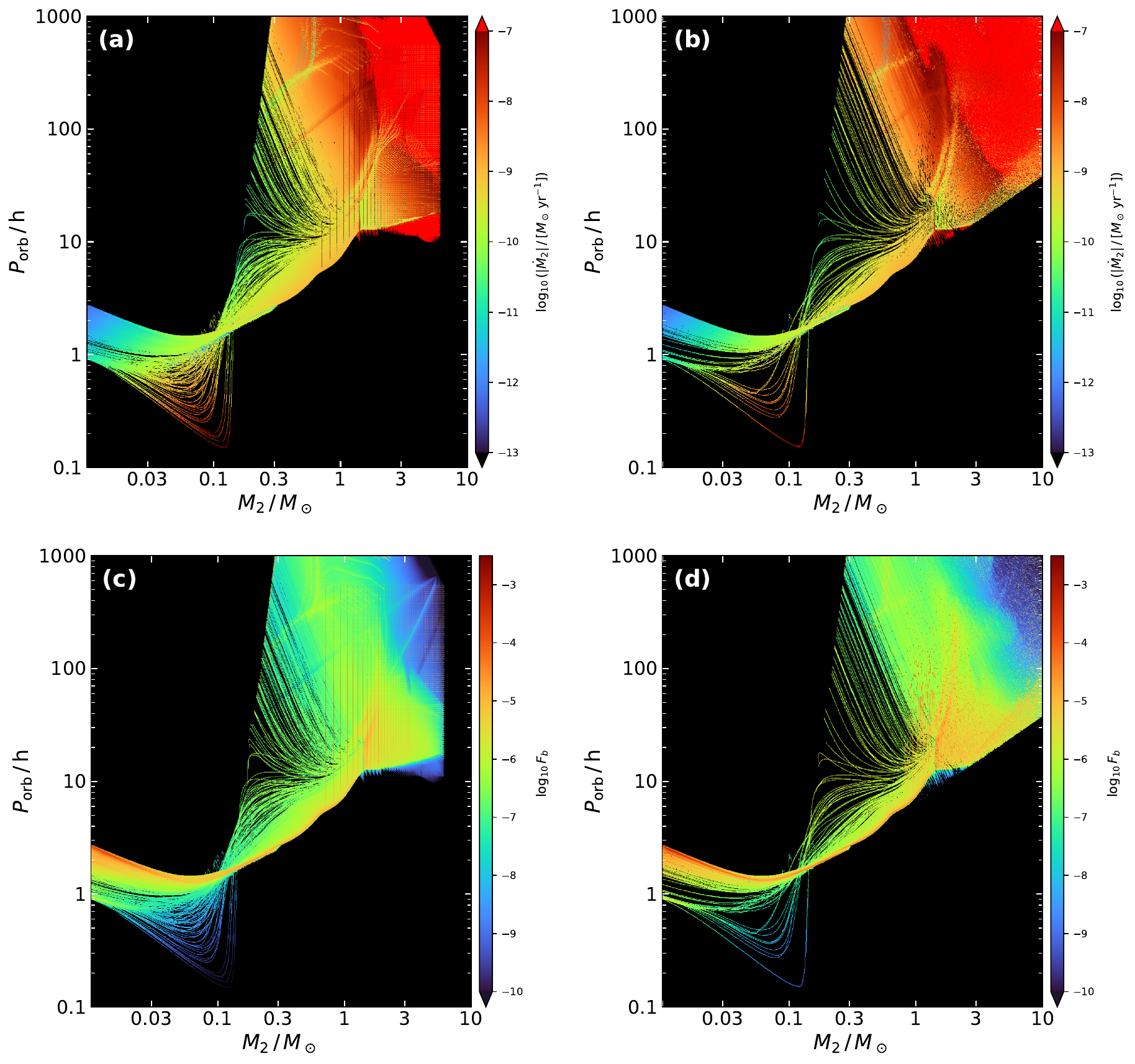}
\caption{Evolutionary tracks in the orbital-period--donor-mass plane.
Panels (a) and (b) are colour-coded by the median
mass-transfer rate in each pixel for the GRID and BPS-seeded
calculations, respectively. Panels (c) and (d) show
\(\log_{10}F_b\), where \(F_b\) is the fraction of the total analysed
evolutionary time accumulated in each displayed pixel, as defined in
equation~\ref{eq:normalised_occupancy}. The normalization is performed
separately for the two initial population samples, so that
\(\sum_{b\in\mathcal{B}}F_b=1\) in each panel.}
\label{fig:Fig03}
\end{figure*}

Fig.~\ref{fig:Fig02} to \ref{fig:Fig07} show the irradiation-free evolutionary
tracks. Fig.~\ref{fig:Fig02} is in the \(P_{\rm orb}\)--\(M_{\rm donor}\) plane. Eighteen observed
systems with dynamical mass constraints are overplotted with their
published donor-mass error bars or one-sided limits. 
The GRID and BPS-seeded calculations trace similar broad
evolutionary loci in this plane. Both include a converging, short-period
branch with low-mass donors, within or close to which many of the observed
systems are located. The calculations also pass through a broader
long-period region containing more massive donors but comparatively
few observed BH--LMXBs occupy this part of the diagram. 
The differences between the two panels, particularly at longer
periods and higher donor masses, reflect the different initial
conditions represented in the GRID and BPS-seeded samples.

XTE J1650--500 and Swift J1727.8--1613 are not plotted because the
adopted source studies do not provide a usable donor mass or mass-ratio
interval. The other 18 systems overlap, or lie close to, the
calculated tracks within their published observational error bars. The
six class-3 systems in Table~\ref{tab:BHList} are excluded because no
donor absorption-line dynamical solution is available for them.
Tracing each observed system and its uncertainty region back to the
corresponding initial parameter space requires a track-by-track inverse
comparison and is not attempted here.

Fig.~\ref{fig:Fig03} shows the same
\(P_{\rm orb}\)--\(M_{\rm donor}\) plane as Fig.~\ref{fig:Fig02} but
now colour-coded by the local mass-transfer rate in panels (a) and (b)
and by the fraction of the total analysed evolutionary time in panels
(c) and (d).
Panels (a) and (b) show that the largest median transfer rates
occur mainly among systems with more massive donors and longer orbital
periods, whereas the converging short-period, low-mass branch generally
has lower transfer rates. These panels locate the rapidly and slowly
transferring regions of the calculations but they do not by themselves
establish whether a particular high-rate segment is driven on a thermal
or nuclear timescale; that interpretation also depends on the donor
structure and mass ratio.

Panels (c) and (d) show where each numerical ensemble
accumulates its total computed evolutionary time. Pixels with larger
\(F_b\) contain a larger fraction of the time summed over all tracks in
that initial population sample. Because the tracks are summed before
normalization, \(F_b\) is not the mean lifetime fraction of an
equal-weight or typical track. Differences between the GRID and
BPS-seeded panels are determined in part by their different initial
conditions and track durations.

Conditional on the formation of incipient black-hole binaries
with low-mass donors, our calculations identify an evolutionary
pathway to a presently unconfirmed class with
\(M_2<0.1\,M_\odot\) and
\(10\,{\rm min}\leq P_{\rm orb}\leq3\,{\rm h}\).\footnote{
But we note the existence of very faint
X-ray transients \citep{DegenaarWijnands2009}. Some contain neutron
stars identified through type-I X-ray bursts, whereas others with
unidentified accretors could harbour black holes. Their inferred
long-term mean accretion rates are low, although uncertain, making
a connection with the systems considered here worth investigating.}
This region is reached by 2,624 tracks across the two irradiation-free
samples, showing that the pathway occurs for multiple adopted
post-black-hole-formation initial configurations. Along much of this
branch, the low mass-transfer rates imply characteristic donor
mass-loss timescales, \(M_2/|\dot M_2|\), of order Gyr, so these systems
can be long-lived. No system satisfying both limits has yet been
securely identified observationally. Because the tracks carry neither
formation nor observational-selection weights, their number cannot
be used to infer a Galactic abundance.

\subsection{Mass-ratio--orbital-period evolution}
\label{subsec:mass_ratio_period}

\begin{figure*}
\centering
\includegraphics[width=0.95\linewidth]{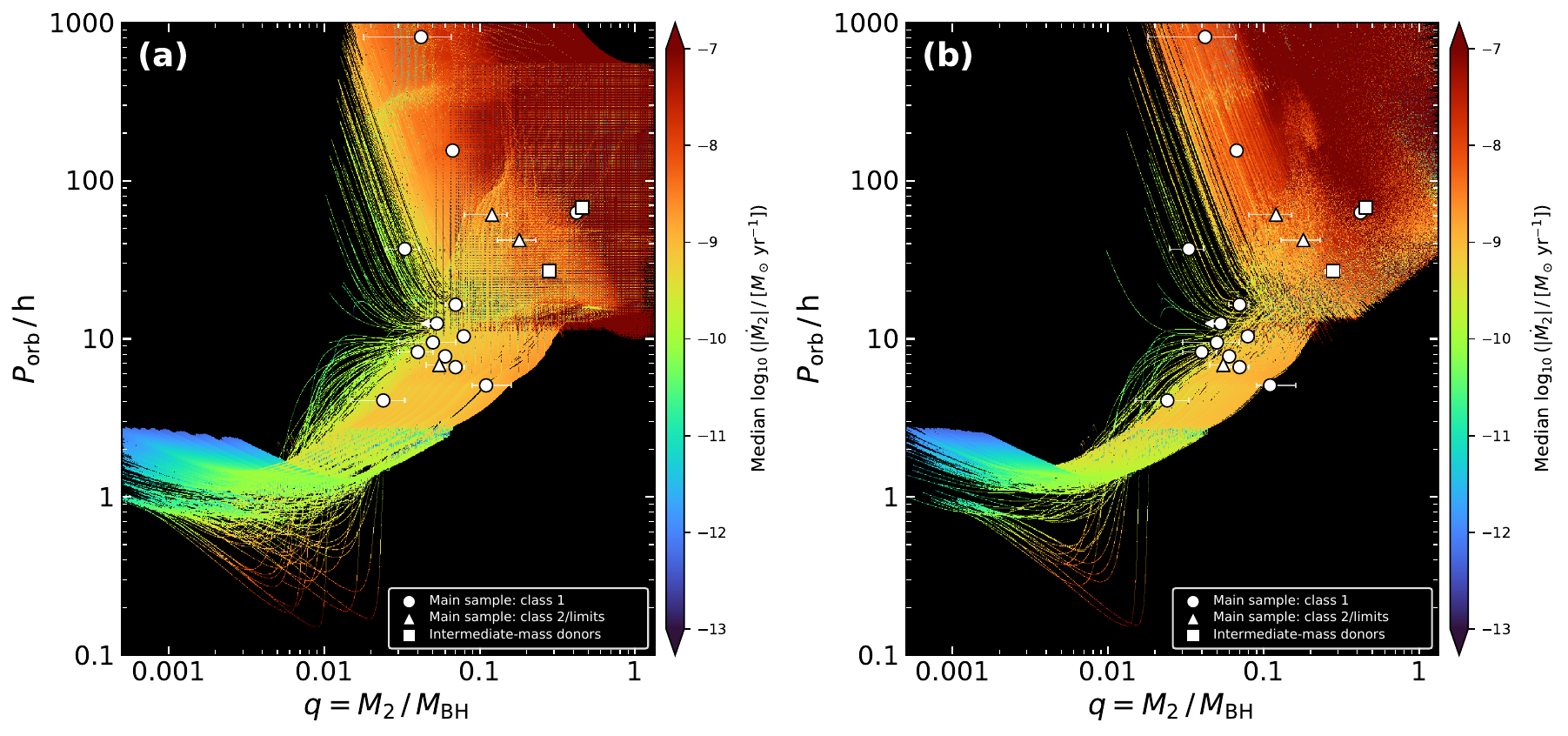}
\caption{Orbital period versus mass ratio,
\(q=M_{\rm donor}/M_{\rm BH}\), for (a) GRID and (b) BPS-seeded models.
Both axes are logarithmically scaled. Colours show the median logarithmic mass transfer rate
\(\log_{10}(|\dot{M}_{\rm donor}|/[M_\odot\,{\rm yr}^{-1}])\)
in each pixel. The observed systems are shown as white circles
(class~1), triangles (class~2), and squares
(the two intermediate-mass donor systems), following the
symbol convention of Fig.~\ref{fig:Fig02}. Error bars show the published
mass-ratio uncertainties. Arrows denote one-sided limits. Class-3
systems are not shown.}
\label{fig:Fig04}
\end{figure*}

Fig.~\ref{fig:Fig04} recasts the evolutionary grid in the
mass-ratio--period plane. Because
\(q=M_{\rm donor}/M_{\rm BH}\), this projection largely reproduces the
morphology already visible in the
\(P_{\rm orb}\)--\(M_{\rm donor}\) plane. The principal additional role
here is to permit a direct comparison with the observed mass ratios and
to show how \(q\) changes as the donor is stripped. We do not use this
plane as an independent diagnostic of mass-transfer stability.

At long orbital periods and \(q\gtrsim0.3\), the tracks correspond mainly
to systems with donors that remain relatively massive compared with their
black-hole accretors. The relatively high mass-transfer rates in
this region indicate rapid donor mass loss but the
\(q\)--\(P_{\rm orb}\) projection alone does not distinguish
thermal-timescale from nuclear-timescale mass transfer.
As mass transfer reduces \(M_{\rm donor}\), the systems migrate leftward
towards lower \(q\). At shorter orbital periods, their secular
evolution becomes increasingly controlled by angular-momentum losses.

At shorter orbital periods
(\(P_{\rm orb}\lesssim5\)--10 hours) and low mass ratios
(\(q\lesssim0.1\)), the models converge towards a narrow sequence
characterised by relatively modest mass-transfer rates.
In Fig.~\ref{fig:Fig03}(c, d), \(F_b\) gives the fraction of the
summed analysed duration of the corresponding model ensemble assigned
to each bin and so quantifies the global dwell time in this
locus. Such a low \(q\) places the donors well below the black-hole
masses so Fig.~\ref{fig:Fig04} is not used to identify a separate
mass-transfer stability boundary. Magnetic braking and gravitational
radiation instead drive the secular evolution along this short-period
branch.

The presence of a low-\(q\), short-period locus in both ensembles shows
that this evolutionary pathway is reached in both
initial population samples. Differences at higher \(q\) and
longer periods demonstrate that the initial sampling still affects
which tracks populate each region.

\subsection{Donor temperature and spectral evolution}
\label{subsec:donor_temperature}

\begin{figure*}
\centering
\includegraphics[width=0.95\linewidth]
{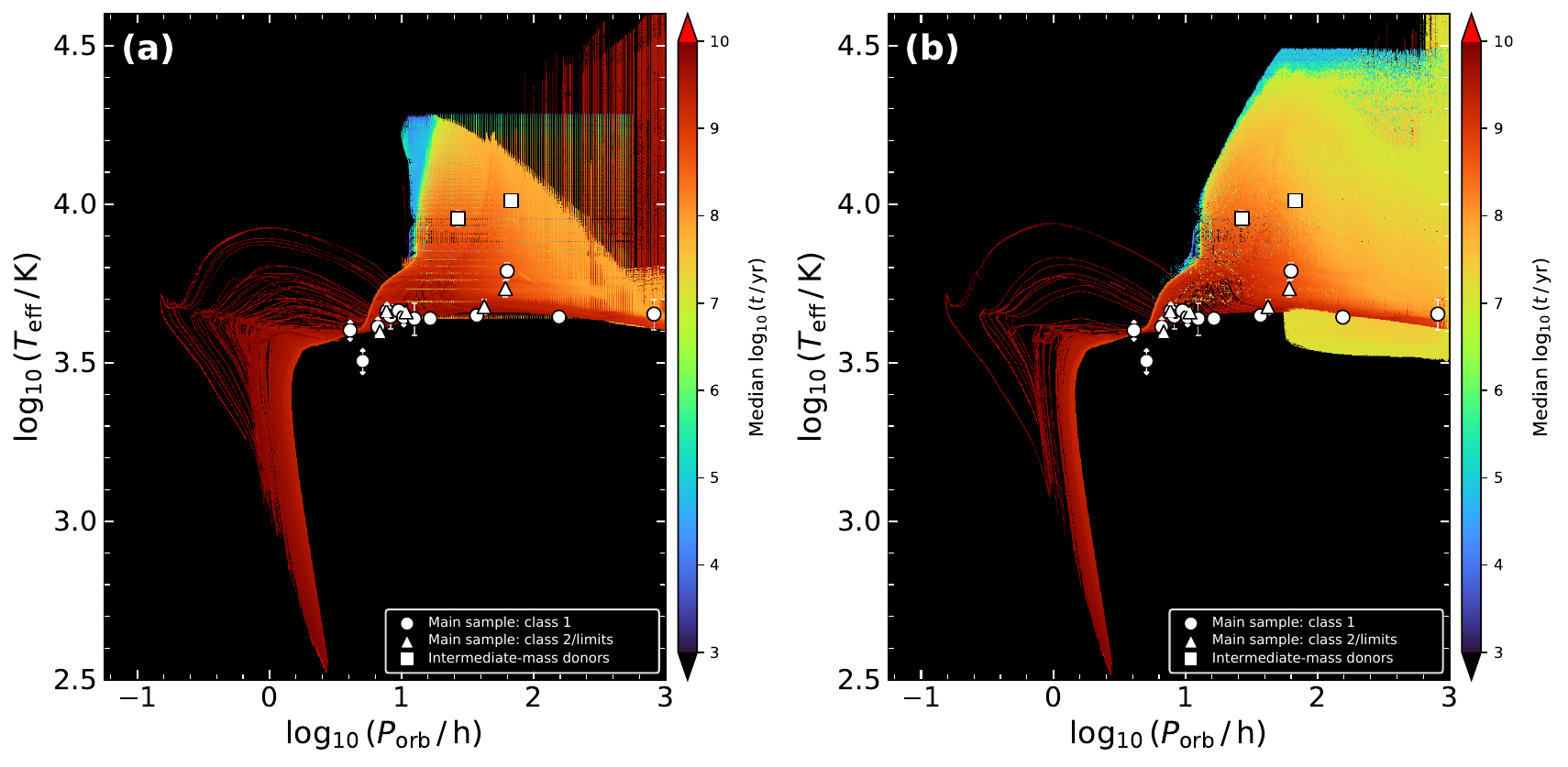}
\caption{Orbital period versus donor effective temperature for
(a) the GRID and (b) the BPS-seeded calculations. Each
\((\log_{10}P_{\rm orb},\log_{10}T_{\rm eff})\) cell is coloured by the
median model age, \(\log_{10}(t/{\rm yr})\), of the models in that cell.
Observed systems are overplotted using the same symbols
as in Fig.~\ref{fig:Fig02}. For systems with an approximate numerical temperature but no
quoted uncertainty, we show an illustrative interval of
\(T_{\rm eff}\pm300\,\mathrm{K}\). Outward-pointing arrowheads
indicate that the uncertainty may be larger. These intervals are
adopted for display and are not published measurement uncertainties.}
\label{fig:Fig05}
\end{figure*}

Fig.~\ref{fig:Fig05} compares the intrinsic donor temperatures
predicted by the irradiation-free calculations with
the observed systems with dynamically determined masses.
The models show a relatively narrow cool-donor
locus at \(3.6\leq \log_{10}(T_{\rm eff}/{\rm K})\leq3.75\), extending from
orbital periods of a few hours to several tens of hours.
Most of the observed systems associated with this locus lie in
its shorter-period part.
Both the GRID and the BPS-seeded calculations reach this region after
substantial donor mass loss and orbital evolution.
This overlap shows that stripped-donor evolution can produce
cool companions at the observed orbital periods. This is the principal
result illustrated in this figure.

At longer periods and higher effective temperatures the
calculations also populate a broader region containing partially
stripped donors of intermediate-mass origin. The median-age colouring
shows when the hot long-period configurations and the cool
short-period loci are reached. 
Figure~\ref{fig:Fig05} shows the spectroscopic and model-adopted
temperatures listed in Table~\ref{tab:BHList}. For five systems with
only a spectral classification in the table, we use illustrative
temperatures inferred from their spectral types. They may be affected by disc veiling,
atmosphere-model assumptions, and irradiation, whereas the
calculations show intrinsic donor effective temperatures. The comparison
therefore tests the broad donor structure and thermal state rather than
the detailed spectrum of an irradiated stellar hemisphere.

\subsection{Time-dependent evolution and orbital bifurcation}
\label{subsec:time_bifurcation}

\begin{figure*}
\centering
\includegraphics[width=0.95\linewidth]{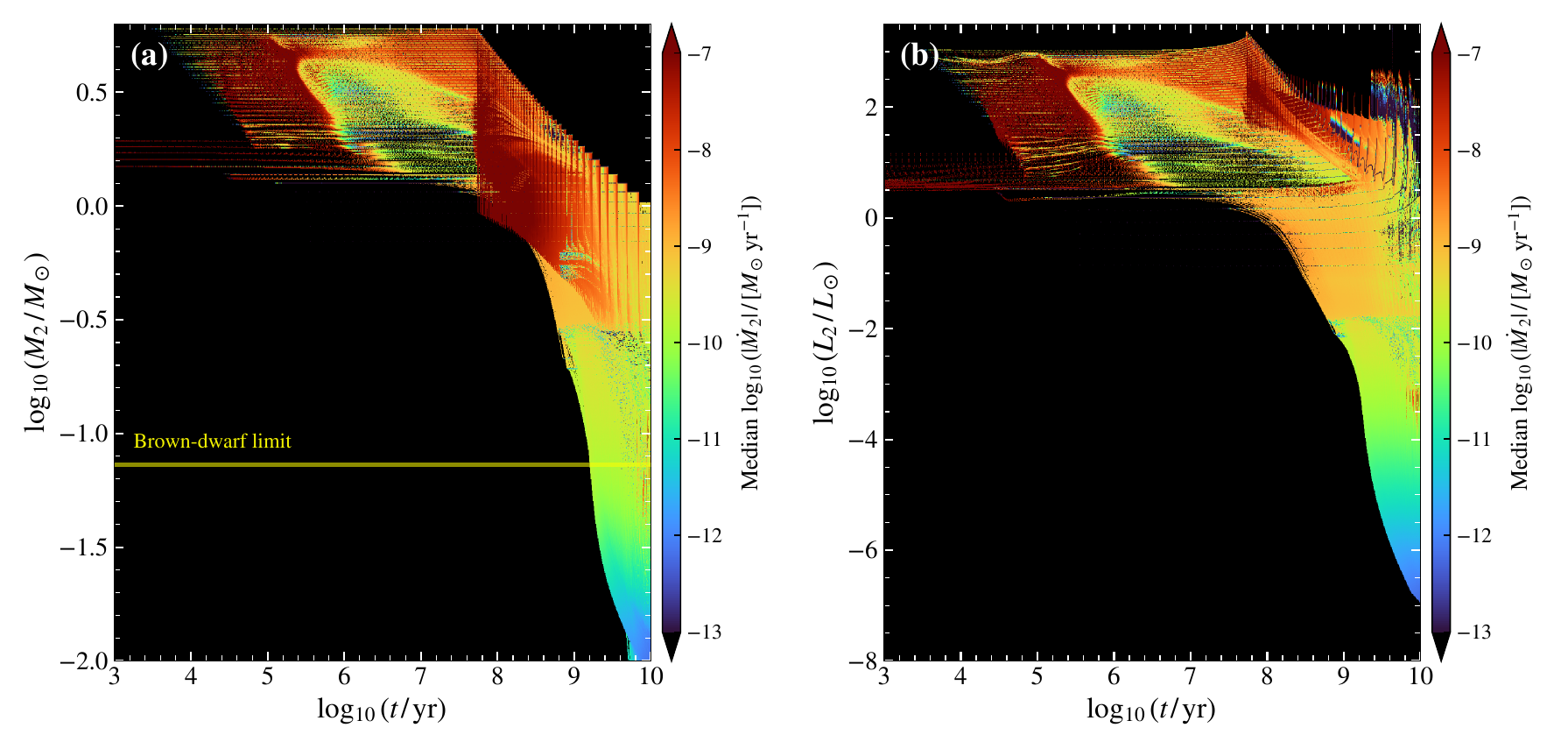}
\caption{Time-domain diagnostics for the irradiation-free GRID
calculations. Panel (a) shows donor mass as a function of model age, and
panel (b) shows donor luminosity as a function of model age.
Each occupied bin is colour-coded by the median
\(\log_{10}(|\dot{M}_2|/[M_\odot\,{\rm yr}^{-1}])\)
in that bin.
In panel (a), the horizontal line marks the hydrogen-burning limit,
\(M_2=0.073\,M_\odot\).}
\label{fig:Fig06}
\end{figure*}

Fig.~\ref{fig:Fig06} follows the evolution of donor mass
and luminosity with model age in the irradiation-free GRID calculations.
Unlike the H--R projection in Fig.~\ref{fig:A1}, this
time-domain view shows when donor stripping and luminosity changes occur
and how they coincide with changes in the mass-transfer rate.

The time-domain maps reveal a clear ordering of the secular evolution.
The strongest mass-transfer rates occur primarily while relatively
massive and luminous donors are being stripped, whereas the later
evolution extends towards lower donor masses, lower luminosities and
generally weaker mass transfer. Some sequences continue below the
hydrogen-burning limit, forming a very low-mass tail at late model
ages. The broad range of donor properties present at a given age
reflects the different initial masses and orbital periods represented
in the GRID rather than a single universal evolutionary sequence.

\begin{figure*}
\centering
\includegraphics[width=0.95\linewidth]
{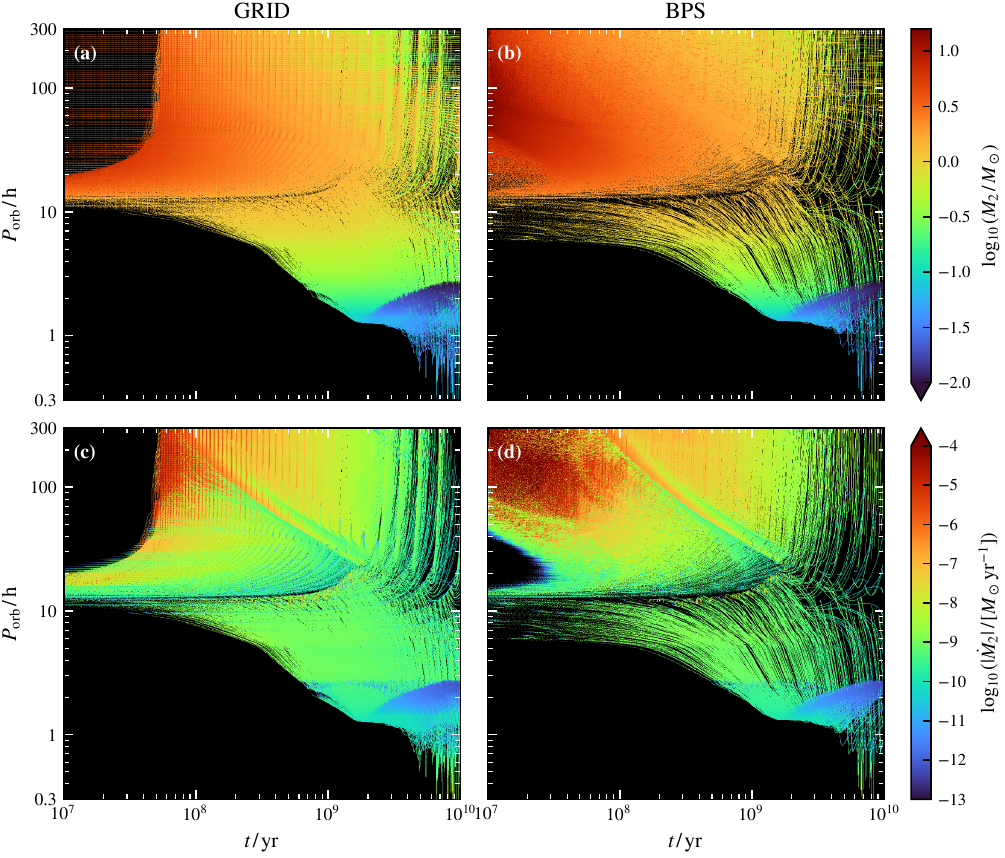}
\caption{Age $t$ -- orbital-period $P_{\rm orb}$ projections of the irradiation-free
reference populations. Panels (a) and (c) show the regular GRID
calculations, while panels (b) and (d) show the BPS-seeded calculations.
The upper panels are colour-coded by the median donor mass,
\(\log_{10}(M_2/M_\odot)\), in each occupied age--period bin.
The lower panels are colour-coded by the median magnitude of the donor
mass-transfer rate,
\(\log_{10}(|\dot{M}_2|/[M_\odot\,{\rm yr}^{-1}])\),
and include only active-transfer samples satisfying
\(|\dot{M}_2|\geq10^{-13}\,M_\odot\,{\rm yr}^{-1}\).
Black cells contain no model samples in the upper panels and no
active-transfer samples in the lower panels.}
\label{fig:Fig07}
\end{figure*}

The age--orbital-period projections in Fig.~\ref{fig:Fig07} illustrate
the coexistence of converging and expanding evolutionary pathways in
both the GRID and BPS-seeded calculations.
In both ensembles, sequences on the converging branch move from
the bifurcation region near \(P_{\rm orb}\sim10\)\,h towards shorter
periods because their donors are stripped to low masses. Sequences on the
expanding branch instead evolve towards orbital periods of tens to
hundreds of hours while retaining more massive donors. The lower panels
show that the strongest mass-transfer rates occur mainly on the
expanding branch, whereas the late short-period branch generally
evolves at lower rates. The same broad bifurcation morphology is present
in both the GRID and BPS-seeded calculations despite their different
initial sampling. Here, coexistence refers to distinct tracks with different initial
conditions occupying both pathways at a given age; it does not imply
that an individual evolutionary track splits into two branches.

\section{Results from the Irradiated Calculations}
\label{sec:irradiation_results}

We next consider the 4,392 calculations that include irradiation
feedback and provide usable donor-mass, orbital-period, and
mass-transfer-rate histories for Fig.~\ref{fig:Fig08}. They sample a
targeted region around the onset of Roche-lobe overflow and the
orbital-period bifurcation, rather than duplicating the full initial
coverage of the irradiation-free GRID and BPS populations.

\subsection{Irradiated mass-transfer morphology}
\label{subsec:irrad_morphology}

\begin{figure*}
\centering
\includegraphics[width=0.7\textwidth]{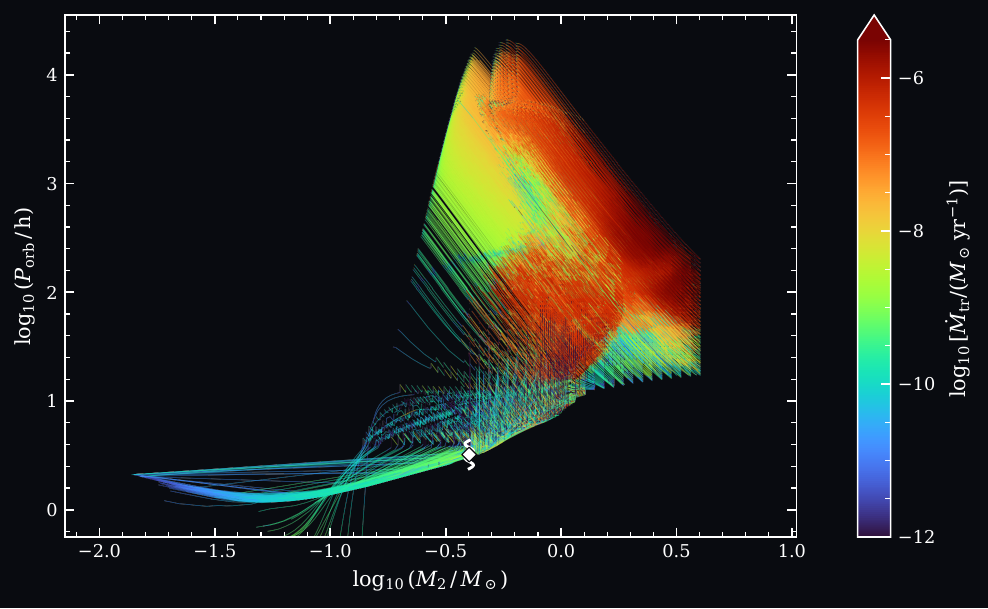}
\caption{Irradiated evolutionary tracks in the
donor-mass--orbital-period plane, combining the initial 5 and
$8\,M_\odot$ black-hole samples. Each track is colour-coded along its
path by the local Roche-lobe mass-transfer rate,
$\log_{10}[|\dot M_{\rm tr}|/(M_\odot\,{\rm yr}^{-1})]$.
The white wavy marker and adjacent diamond indicate approximately
where the adopted prescription reduces $\eta_{\rm irr}$ from 0.05 to
$10^{-5}$, effectively switching off irradiation for tracks below
and to the left to keep the calculations tractable. The full
criterion is timestep dependent, so the marker is a visual guide
rather than a sharp physical boundary. Overplotted tracks are not
weighted by evolutionary dwell time or binary-formation
probability.}
\label{fig:Fig08}
\end{figure*}

Fig.~\ref{fig:Fig08} combines the irradiated evolutionary tracks
calculated with initial black-hole masses of 5 and $8\,M_\odot$ in the
donor-mass--orbital-period plane. The two subsets, defined by their
different initial black-hole masses, show the same large-scale
morphology. Each track is colour-coded along its path by the local
Roche-lobe mass-transfer rate, $\dot M_{\rm tr}$. The white wavy marker
and adjacent diamond show approximately where the adopted prescription
reduces $\eta_{\rm irr}$ from 0.05 to $10^{-5}$. For model states below
and to the left of the marker, irradiation is effectively switched off
to keep the calculations tractable. The reduction criterion also
depends on the timestep and is evaluated at each model, so the marker
is a visual guide rather than a sharp boundary or a permanent switch.

The mass-transfer rate tends to be high around
\(M_2\simeq1\,M_\odot\) and
\(P_{\rm orb}\simeq50\,\mathrm{h}\) on the expanding branch
in Fig.~\ref{fig:Fig08}.
The rates are somewhat lower in the same region of the
irradiation-free maps in Fig.~\ref{fig:Fig03}(a,b),
suggesting that mass transfer there is enhanced by irradiation.
Within the irradiated sample, the converging low-mass branch
has lower secular mass-transfer rates than the high-transfer-rate
region on the expanding branch described above, and evolves towards
the period minimum.

The converging irradiated tracks extend to very low donor masses
and short orbital periods. Their converged portions continue towards the short-period tail
under the adopted state-dependent reduction in irradiation
efficiency; they do not test sustained irradiation at the nominal
efficiency along that entire branch. Tracks that stop for numerical reasons are not
interpreted as physical terminations of that sequence.

The irradiated tracks in Fig.~\ref{fig:Fig08} reproduce the
broad morphology of the irradiation-free maps in
Figs~\ref{fig:Fig02} and~\ref{fig:Fig03}, including the converging
short-period and expanding long-period branches. This agreement
supports using the much larger irradiation-free samples to map the
broad secular evolutionary pathways, while examining the
irradiation-driven variability of mass transfer separately.
The lowest-mass, short-period portions of the irradiated tracks
are evolved with strongly reduced irradiation efficiency. 
Differences in track density between the irradiated sample
in Fig.~\ref{fig:Fig08} and the irradiation-free samples in
Figs~\ref{fig:Fig02} and~\ref{fig:Fig03} cannot be attributed to
irradiation alone, because these samples differ in their initial
sampling and mass-retention prescriptions.

\subsection{Detailed Look at the Irradiation Cycles}
\label{sec:detailed_irradiation}
\label{subsec:matched_irrad}

\begin{figure*}
\centering
\includegraphics[width=0.85\textwidth]{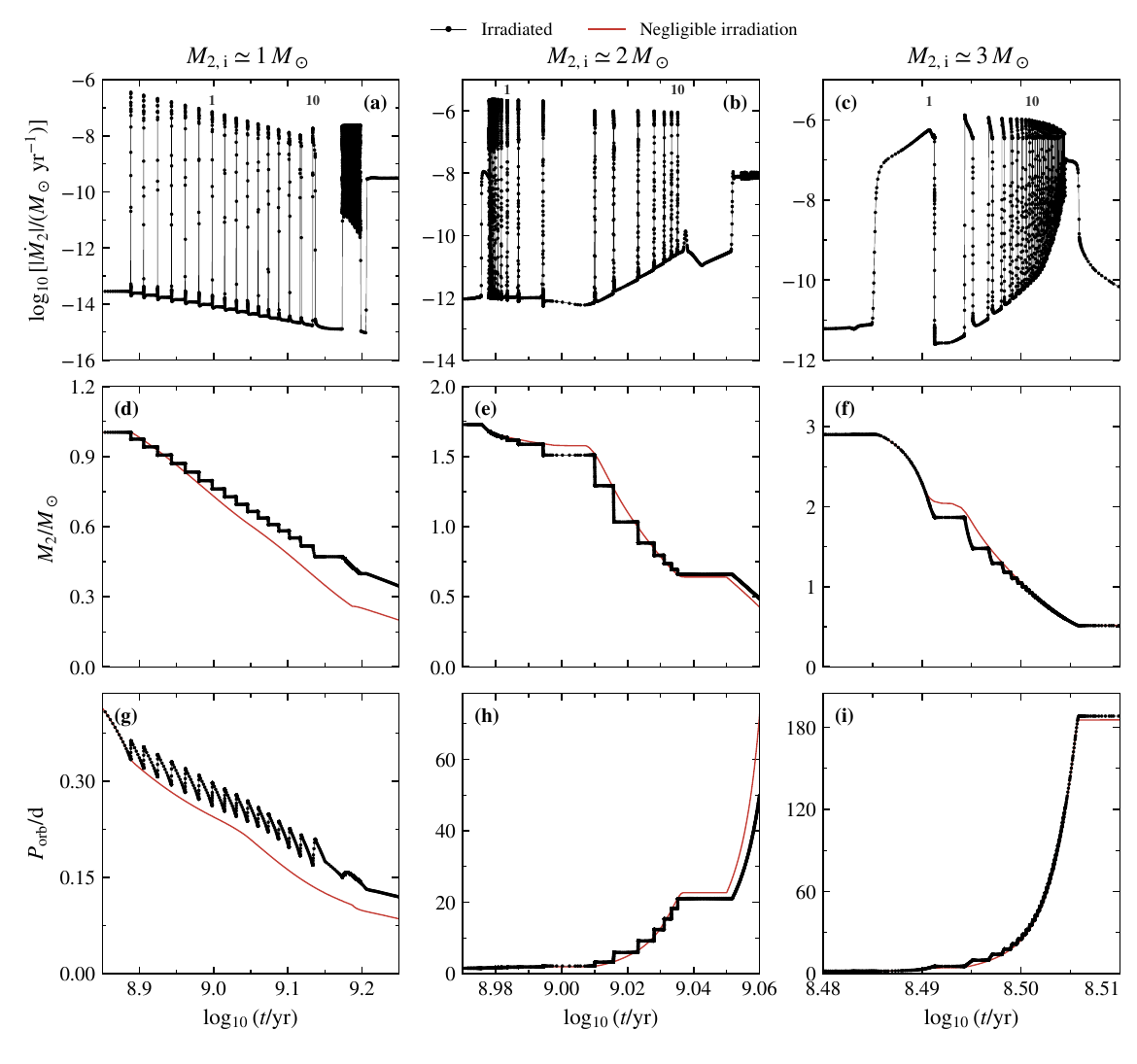}
\caption{Examples of irradiation cycles for three binaries with
an initial black-hole mass of \(8\,M_\odot\). 
From left to right, the initial donor masses
are 1.00, 2.03 and \(3.00\,M_\odot\), and the initial orbital
periods are 17.5, 24.8 and 39.9\,h, respectively.
The rows show the total donor mass-loss rate, donor mass and
orbital period as functions of logarithmic age. The black points
mark all stored states in the selected intervals, and thin lines
connect successive points. The donor mass-loss rate includes
both Roche-lobe overflow and wind loss; it is not the rate
accreted by the black hole.
In the lower two rows, red curves show the corresponding
calculations with the same initial conditions, mass-retention
prescription and other input controls, but with the irradiation
efficiency fixed at \(\eta_{\rm irr}=10^{-10}\), making irradiation
negligible. Both calculations are plotted at their absolute model
ages, without a time shift. The panels show selected evolutionary
intervals rather than complete lifetimes, and use different axis
limits between columns. Labels 1 and 10 identify the first and last
selected episodes tabulated in Table~\ref{tab:irradiation_episodes}.}
\label{fig:irradiation}
\label{fig:Fig09}
\end{figure*}

In Fig.~\ref{fig:irradiation} we show three illustrative BH-binary evolutions with irradiation for three different initial mass donor stars.  In each case, we plot $\dot{M}_{\rm 2}$, $M_{\rm 2}$, and $P$ vs.~evolution time. What we see in all cases are irradiation-driven mass-transfer cycles where $\dot{M}_{\rm 2}$ varies by several orders of magnitude.  At the same time the evolutions of $M_{\rm 2}$, and $P$ vary by relatively small steps that are mostly monotonic in direction.  This bolsters the validity of our use of the non-irradiated models to represent the evolution of BH binaries in the $P-M_{\rm 2}$ plane (see Sect.~\ref{sec:secular_evolution} for a further discussion).

The sharp spikes in $\dot{M}_{\rm 2}$
can often reach quite high values approaching or even exceeding
$10^{-6}$ M$_\odot$ yr$^{-1}$.
For the selected sharp events in the initially 2 and
$3\,M_\odot$ examples, the durations are approximately
$5\times10^4$--$6\times10^5$ yr and the donor-mass losses are
0.017--$0.386\,M_\odot$.
The lower-rate events in the initially $1\,M_\odot$ example
are substantially longer, as quantified below.
We also often find much longer duration irradiation-driven
mass-loss events with $\dot M_2$ in a much lower range of
$\sim$$10^{-8}-10^{-7}$ M$_\odot$ yr$^{-1}$, that can precede
and follow the sharper spikes of mass transfer.
In these cases, especially for such longer events preceding
the string of sharp spikes, the net amount of mass loss can
be rather substantial, and can remove as much as 30\% to 50\%
of the total mass of the donor star.

To make some of the above numerical estimates more concrete, we have constructed a table of quantitative values for 10 of the irradiation spikes for each initial donor mass shown in Fig.~\ref{fig:irradiation}.  In particular, we tabulate (i) the duration of the event, $\log (\Delta t)$, (ii) the duration of the event divided by the time interval to the previous outburst, $\Delta t/t_{\rm prev}$, (iii) the mean logarithmic donor mass-loss rate during the outburst, $\bar\ell$, and (iv) the total mass lost during the event, $\Delta M_2$.  The numerical values for the outburst properties are summarized in Table \ref{tbl:properties}.

In the initially $1\,M_\odot$ example,
the selected events broaden from 0.69 to 10.45 Myr and
the duration fraction rises from 0.017 to 0.20, while
the mean logarithmic mass-loss rate decreases from
$-7.30$ to $-8.38$.
The duration fractions span 0.008--0.031 and 0.092--0.228
in the initially 2 and $3\,M_\odot$ examples, respectively. The first broad event in the $3\,M_\odot$ example removes $1.03\,M_\odot$, about 35 per cent of its initial donor mass; the remaining selected events across the three examples remove 0.017--$0.386\,M_\odot$ each. These event measurements do not represent lifetime duty cycles. Since the initial periods and evolutionary phases also differ, the differences between these examples cannot be attributed to donor mass alone.

Finally, in this regard, we note that these effective `on-off' times in the irradiation cycle could have significant implications for estimating BH-binary formation rates from the numbers of existing systems.  In particular, there are long intervals when these systems are effectively in an `off' state, in which case there could be many times more latent (i.e., dormant) systems than the ones we observe.  These selected events alone do not determine
a population-wide `on-off' duty cycle.

\subsection{Secular Evolution Remains the Same}
\label{sec:secular_evolution}

To see how the illustrative irradiation outburst events in Fig.~\ref{fig:irradiation} compare with the corresponding non-irradiated cases, we first identified the initial parameters used to run each case in Fig.~\ref{fig:irradiation}, e.g., $P_{\rm init}$, $M_{\rm 2,init}$, and $M_{\rm BH,init}$. We then made MESA runs for the same initial systems,
but with essentially no irradiation.  The evolutionary results for $M_{\rm 2}$ and $P$ are overplotted on the corresponding curves in Fig.~\ref{fig:irradiation} as red curves. Here we see that the red curves smoothly follow the trends of the steps in the irradiated curves.  The disparate timescales in the three cases are closely followed by the red curves.  The matches in absolute values of $M_2$ and $P$ (especially in the case of initially 1 M$_\odot$ donor) are adequate, but not perfect. However, one would hardly expect two such very different physical evolutions to simply replicate each other's mean values with perfect fidelity.

From this study, we can draw the conclusion that we were indeed
justified in leaving out irradiation when we studied large numbers
of tracks in the $P-M_2$ plane. The irradiation-free reference samples
contain more than an order of magnitude more tracks than the
irradiated sample.

To summarize, in the irradiated models $\dot M_2$ varies by several
orders of magnitude with time, but the steps in $P$ and $M_2$ are
relatively small. The latter are quite reasonably well matched by
the smooth variations in the non-irradiated models.

\begin{table*}
\centering
\caption{Illustrative Irradiation Driven $\dot M$ Episodes}
\label{tab:irradiation_episodes}
\label{tbl:properties}
\setlength{\tabcolsep}{3.0pt}
\small
\begin{tabular}{r rrrr rrrr rrrr}
\toprule
 & \multicolumn{4}{c}{$M_{2,\mathrm{i}}=1.00374\,M_\odot$}
 & \multicolumn{4}{c}{$M_{2,\mathrm{i}}=2.03203\,M_\odot$}
 & \multicolumn{4}{c}{$M_{2,\mathrm{i}}=2.99900\,M_\odot$} \\
\cmidrule(lr){2-5}\cmidrule(lr){6-9}\cmidrule(lr){10-13}
No. & $\log\Delta t$ & $f_{\rm prev}$ & $\bar\ell$ & $\Delta M$
    & $\log\Delta t$ & $f_{\rm prev}$ & $\bar\ell$ & $\Delta M$
    & $\log\Delta t$ & $f_{\rm prev}$ & $\bar\ell$ & $\Delta M$ \\
\midrule
1 & 5.84 & 0.017 & -7.30 & 0.035 & 4.89 & 0.019 & -6.77 & 0.017 & 6.62$^{\ddagger}$ & 0.228 & -6.72 & 1.034 \\
2 & 5.91 & 0.021 & -7.38 & 0.034 & 5.01 & 0.014 & -6.63 & 0.030 & 5.77 & 0.092 & -6.20 & 0.386 \\
3 & 6.00 & 0.026 & -7.49 & 0.032 & 5.15 & 0.008 & -6.32 & 0.077 & 5.45 & 0.164 & -6.21 & 0.183 \\
4 & 6.07 & 0.031 & -7.58 & 0.030 & 5.55 & 0.010 & -6.23 & 0.218 & 5.25 & 0.180 & -6.22 & 0.111 \\
5 & 6.15 & 0.038 & -7.69 & 0.028 & 5.62 & 0.031 & -6.22 & 0.259 & 5.10 & 0.196 & -6.22 & 0.077 \\
6 & 6.24 & 0.048 & -7.80 & 0.027 & 5.34 & 0.012 & -6.20 & 0.147 & 4.98 & 0.206 & -6.22 & 0.059 \\
7 & 6.35 & 0.057 & -7.91 & 0.027 & 5.13 & 0.011 & -6.21 & 0.086 & 4.88 & 0.212 & -6.22 & 0.047 \\
8 & 6.51 & 0.079 & -8.03 & 0.030 & 4.96$^{\dagger}$ & 0.012 & -6.24 & 0.055 & 4.80 & 0.216 & -6.22 & 0.039 \\
9 & 6.70 & 0.113 & -8.16 & 0.035 & 4.83$^{\dagger}$ & 0.012 & -6.25 & 0.040 & 4.75 & 0.227 & -6.23 & 0.034 \\
10 & 7.02 & 0.200 & -8.38 & 0.046 & 4.73$^{\dagger}$ & 0.011 & -6.26 & 0.031 & 4.69 & 0.219 & -6.24 & 0.029 \\
\bottomrule
\end{tabular}%
\par\smallskip
\begin{minipage}{\textwidth}
\footnotesize
All logarithms are base ten; $\Delta t$ is in yr and $\Delta M$ in $M_\odot$.
The event boundaries are given by the times
of the steep shoulders of the rise and fall.
Here $f_{\rm prev}=\Delta t/(t_{\rm on}-t_{{\rm on},{\rm prev}})$
uses the onset of the actual preceding event, including events outside
the displayed interval.
The mean logarithmic donor mass-loss rate is
$\bar\ell=\Delta t^{-1}\int_{t_{\rm on}}^{t_{\rm off}}
\log_{10}[|\dot M_2|/(M_\odot\,\mathrm{yr}^{-1})]\,\mathrm{d}t$;
it is not the logarithm of the arithmetic mean rate.
The tabulated mass loss is $M_2(t_{\rm on})-M_2(t_{\rm off})$.
The tabulated values are approximate
measurements of selected events; the $f_{\rm prev}$ values
are not lifetime duty cycles.
$^{\dagger}$The combined timing uncertainty
of the rising and falling edges is 15--18 per cent
of the event duration.
$^{\ddagger}$The first event is broad and has a gradual rising shoulder;
its duration and mean depend on the stated onset convention.
\end{minipage}
\end{table*}

\section{Gravitational-wave Strain Along the Irradiation-free Tracks}
\label{sec:compact_lisa}

We use the irradiation-free GRID and BPS samples to examine
which evolutionary states reached under their adopted
prescriptions could produce detectable low-frequency
gravitational-wave signals. 
We evaluate approximately \(6.70\times10^7\) valid stored
evolutionary states from 78,795 tracks, corresponding to about
850 states per track on average, rather than selecting only the
last stored model or the minimum-period state.
No endpoint-based preselection is applied. 
Each stored state is treated as a
quasi-circular binary radiating predominantly in the quadrupole
harmonic.

At each stored state, we define
\(f_{\rm GW}=2/P_{\rm orb}\) as the gravitational-wave frequency of
the dominant quadrupole (\(n=2\)) harmonic of the assumed circular
binary, and calculate the strain from the instantaneous black-hole and
donor masses. The nominal characteristic strain and signal-to-noise ratio are evaluated
at \(d=1\,\mathrm{kpc}\) for a four-year LISA observation.
To account for randomly oriented orbital planes, we
average over an isotropic distribution of binary inclinations. This
gives a factor of \(4/5\) in the squared strain, equivalent to
multiplying the strain amplitude by \(\sqrt{4/5}\). We adopt the
sky-averaged instrumental noise and four-year Galactic confusion
foreground of \citet{Robson2019}.

The chirp mass and strain amplitude at distance \(d\) are
\begin{align}
\mathcal{M}_{\rm c}
&=\frac{(M_{\rm BH}M_2)^{3/5}}{(M_{\rm BH}+M_2)^{1/5}},
\\
h_0
&=\frac{4(G\mathcal{M}_{\rm c})^{5/3}
(\pi f_{\rm GW})^{2/3}}{c^4d}.
\label{eq:lisa_amplitude}
\end{align}

For \(T_{\rm obs}=4\,\mathrm{yr}\), we denote the total one-sided,
sky-averaged strain-noise power spectral density by \(S_n(f)\), with
units of \(\mathrm{Hz}^{-1}\). It is the sum of the instrumental
contribution and the four-year Galactic confusion foreground,
\[
S_n(f)=S_{\rm inst}(f)+S_{\rm conf}(f;T_{\rm obs}=4\,\mathrm{yr}),
\]
following \citet{Robson2019}. Its value at the source frequency is
therefore \(S_n(f_{\rm GW})\). The characteristic strain,
characteristic noise amplitude, and signal-to-noise ratio are
\begin{align}
h_{\rm c}
&=h_0\left(\frac{4}{5}f_{\rm GW}T_{\rm obs}\right)^{1/2},
&
h_n
&=\left[f_{\rm GW}S_n(f_{\rm GW})\right]^{1/2},
\\
\rho
&=\frac{h_{\rm c}}{h_n}
=h_0\left[\frac{(4/5)T_{\rm obs}}
{S_n(f_{\rm GW})}\right]^{1/2}.
\label{eq:lisa_snr}
\end{align}
Here \(h_n\) is the dimensionless characteristic noise amplitude,
whereas \(S_n\) is the corresponding strain-noise power spectral
density. The factor \(4/5\) applies to the squared strain and averages
the two gravitational-wave polarizations over an isotropic
distribution of binary inclinations.

The quantity \(\rho\) measures the strength of the predicted
signal relative to the adopted LISA noise and is therefore the
detectability statistic used in this analysis. We adopt
\(\rho\geq7\) as the nominal four-year detection threshold.

\begin{figure*}
\centering
\includegraphics[width=0.95\linewidth]
{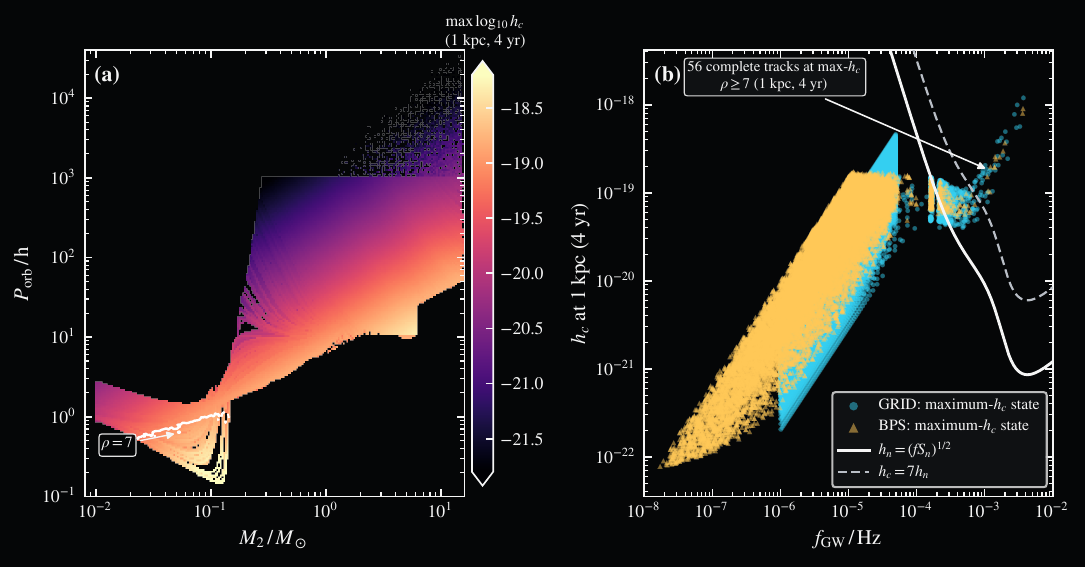}
\caption{Four-year LISA diagnostic for the irradiation-free
GRID and BPS evolutionary calculations at 1 kpc. Panel (a) shows, in each
donor-mass--orbital-period bin, the maximum characteristic strain \(h_c\)
attained by any stored model state and the white contour marks
\(\rho=7\). Panel (b) shows the maximum \(h_c\) state of each complete GRID
track (cyan circles) and complete BPS track
(amber triangles), together with the
sky-averaged four-year LISA characteristic-noise curve
\(h_n=(f_{\rm GW}S_n)^{1/2}\) and the \(\rho=7\) threshold \(h_{\rm c}=7h_n\).
All valid stored evolutionary states were evaluated. 
The ten BPS late-evolution suffixes  are retained in panel (a)
but omitted from the complete-history panel (b). The figure is a
reachability diagnostic and does not include binary-formation weights or
observational-selection effects.}
\label{fig:Fig10}
\end{figure*}

Panel (a) of Fig.~\ref{fig:Fig10} bins the combined GRID and
BPS states in the \(M_2\)--\(P_{\rm orb}\) plane and records the
largest \(h_{\rm c}\) reached in each occupied bin. Taking the maximum
rather than the number of stored points prevents the colour scale from
being interpreted as a dwell-time or population density. The white
contour marks bins in which at least one stored state attains the
nominal threshold \(\rho=7\) at 1 kpc. Panel (b) gives each track equal
visual representation by plotting only the state at which that track
reaches its largest \(h_{\rm c}\). 
The ten BPS tracks that contain only late-evolution suffixes
contribute valid stored states to panel (a) but are omitted from panel
(b) and from statistics requiring a complete stored history.

Of the 78,795 usable tracks, 2,642 attain \(\rho\geq1\) at
1 kpc at some stored state and 60 attain the nominal threshold
\(\rho\geq7\); the latter comprise 47 GRID and 13 BPS tracks. At the
maximum strain states plotted in panel (b) 56 complete tracks remain
above \(\rho=7\). The largest characteristic strain is
\(h_{\rm c}=1.19\times10^{-18}\). Among the 78,785 complete histories,
the maximum-\(h_{\rm c}\) state coincides with the minimum stored
orbital period for 46,076 tracks (58.5 per cent), demonstrating that a
minimum period only selection is not equivalent to evaluating the
strain along the full evolution. 
For the same source parameters and four-year noise model,
\(\rho\propto d^{-1}\). At assumed common distances of 3 and
8 kpc, respectively, 31 tracks  and
21 tracks attain \(\rho\geq7\) at some
stored state. These are unweighted counts of calculated tracks
at the stated distances, not predicted Galactic detection yields.

A total of 2,624 tracks, including 2,093 GRID and 531 BPS
tracks, pass through
\(M_2<0.1\,M_\odot\) and
\(10\,{\rm min}\leq P_{\rm orb}\leq3\,{\rm h}\). This identifies a
specific low-mass, short-period region in which electromagnetic
searches and LISA cross-matching may find previously unrecognized
black-hole binaries. These are unweighted model-track counts and must
not be converted into a Galactic abundance or detection yield without
formation weights, residence times, and survey selection effects.

\section[Discussion]{{\textbf{Discussion}}}

In this work, we have carried out detailed black-hole binary evolutions
with \textsc{MESA}, beginning after the black hole had formed. The
irradiation-free analysis contains approximately 79,000 usable
tracks drawn from two complementary sets of post-formation BH--donor
binaries: a finely sampled regular GRID and an auxiliary BPS-seeded
population. A separate irradiated sample contains 4,392 usable
evolutionary tracks.

The results are presented as evolutionary tracks in the
\(P_{\rm orb}\)--\(M_2\), \(P_{\rm orb}\)--mass-ratio,
\(L\)--\(T_{\rm eff}\), and \(T_{\rm eff}\)--\(P_{\rm orb}\) planes,
together with the evolution of donor properties as a function of time.
Across these evolutionary-track figures, colour encodes
quantities including the mass-transfer rate, relative dwell time, age,
gravitational-wave strain, black-hole mass growth, and donor surface
composition. Each computed track is counted once, without binary-formation
or observational-selection weights. The samples therefore identify
reachable evolutionary states, not Galactic formation rates or
predicted source counts.

Since much of the evolutionary morphology is most readily understood in
the $P_{\rm orb}$--$M_2$ plane, we summarize the calculations in terms of
three broad evolutionary categories.

\subsection[Giant branch]{Giant branch}
These are systems in which the dominant driver of mass transfer is the
nuclear evolution of the donor star.  Their orbital periods generally
increase as mass is transferred from the lower-mass donor to the
higher-mass black hole, and most occupy periods
$P_{\rm orb}\gtrsim12\,\mathrm{h}$.
The irradiation-free reference samples and
the irradiated production calculations use different
mass-retention prescriptions.
The amount of orbital widening depends on the adopted retention
fraction and on systemic mass and angular-momentum loss.
The long-period observed systems GS~2023+338 and GRS~1915+105, with
periods of approximately 6.5 and 34~d, respectively, are natural members
of this expanding branch.  Other observed systems at periods above approximately
20~h may also evolve onto it, although their precise evolutionary states
depend on the donor structure when Roche-lobe overflow begins.
Systems on this branch commonly reach substantial secular mass-transfer
rates and can therefore become conspicuous X-ray sources, subject to the
accretion-disc state and observational duty cycle.

\subsection[Detached black-hole--white-dwarf descendants]{Detached black-hole--white-dwarf descendants}
There should also be a reservoir of systems in which giant-branch mass
transfer has ended, leaving a black hole with a white-dwarf companion.
For donors initially below approximately \(2.2\,M_\odot\), the
core-mass--radius relation is expected to produce the familiar relation
between final orbital period and white-dwarf mass
\citep{Rappaport1995,TaurisSavonije1999,Kalomeni2016}, largely independent
of the accretor mass. More massive donors need not obey this relation and
may leave CO- or ONe-rich remnants on a distinct branch. Such detached
systems could be identified through radial-velocity monitoring of
apparently single white dwarfs or, at sufficiently long periods, through
astrometric orbits with large inferred mass functions.
A clear test of this formation channel is that the resulting
BH--WD binaries should have nearly circular orbits, unlike ordinary
wide binaries that have not undergone mass-transfer-driven
circularisation. Such systems would be analogous to massive
white-dwarf--neutron-star binaries such as PSR~J1614--2230
\citep{Demorest2010,Lin2011}.

Some irradiation-free sequences reach the adopted stopping limit of
\(P_{\rm orb}=1000\,\mathrm{h}\) while their stripped donors remain
extended. We therefore do not infer a terminal CO/ONe-white-dwarf
distribution or fit a final-period--white-dwarf-mass relation from
these endpoints.

To explore the subsequent remnant evolution, we examined a separate
set of 25 irradiation-free calculations with MESA r26.4.1, an initial
black-hole mass of $8\,M_\odot$, donor masses of 2, 3, 4, 5 and
$6\,M_\odot$, and initial periods of 2, 3, 5, 10 and 20~d.
These calculations omit the 1000-h period limit and adopt $Z=0.02$
and non-conservative transfer with $\beta_{\rm MT}=0.7$ and
Eddington-limited accretion. They are separate from the 78,795-track
irradiation-free reference sample.

We identify compact, detached remnants by requiring
$R_2\leq0.2\,R_\odot$,
$\log_{10}[g/({\rm cm\,s^{-2}})]\geq5.5$,
$R_2\leq0.99R_{\rm L}$ and
$|\dot M_{\rm tr}|\leq10^{-12}\,M_\odot\,{\rm yr}^{-1}$
throughout a final stored interval lasting at least $10^5$~yr,
where $R_{\rm L}$ is the donor's Roche-lobe radius.
In the analysed histories, 17 sequences satisfy these criteria with
CO-rich cores and five with He-rich cores; the remaining three do not
yet have verified compact endpoints.
The compact CO-core remnants have masses of
$0.501$--$0.798\,M_\odot$ and orbital periods of 233--1317~d.

A rough description of the remnant-mass trend is
\begin{equation}
\frac{M_{\rm rem}}{M_\odot}
\simeq 0.166 + 0.116\,\frac{M_{2,\rm init}}{M_\odot},
\label{eq:pilot_remnant_mass}
\end{equation}
where $M_{\rm rem}$ is the mass of the compact remnant.
For the 17 verified CO-core remnants, with initial donor masses of
$3$--$6\,M_\odot$, the rms departure from this approximation is
$0.045\,M_\odot$. This is an approximate trend within this grid;
the remnant mass also depends on the initial orbital period.

For these CO-core remnants, the orbital period varies by less than
0.12 per cent over the final compact interval. Some retain appreciable
helium burning, so they are distinguished from white dwarfs already
established on a cooling sequence. Among the verified compact remnants,
a longer initial period leads to a longer remnant-stage period at fixed
initial donor mass.

\subsection[Compact branch]{Compact branch}
The compact branch evolves towards shorter
periods and lower donor masses under angular-momentum loss.
The bifurcation boundary and the evolutionary timescales depend
on the donor's structure at Roche-lobe contact, the adopted
magnetic-braking prescription, and the assumed magnitude of orbital
angular momentum dissipation due to systemic mass loss.
These predictions remain conditional on the formation of the
low-mass-donor binaries included in our initial samples.

On this branch, the sequences reach minimum
orbital periods of order 1--1.5~h and substellar donor masses,
overlapping the approximately \(0.02\)--\(0.08\,M_\odot\) interval
\citep[see also][]{DengLi2024}. These are stripped donors that retain
hydrogen, rather than primordial brown dwarfs. After the period
minimum, orbital expansion is expected as the donor expands in
response to further mass loss.

Across the two irradiation-free samples,
2,624 tracks pass through \(M_2<0.1\,M_\odot\) and
\(10\,\mathrm{min}\leq P_{\rm orb}\leq3\,\mathrm{h}\).
This is an unweighted reachability count for the adopted initial
populations, not a predicted Galactic abundance. The relative-dwell-time
maps show that parts of this low-mass, short-period corridor are
occupied for relatively long intervals within the calculated
sequences.

\subsection[Transition systems]{Transition systems}
These systems initially move towards, or linger near, the expanding
branch but subsequently turn towards ultrashort periods. Their donors
have been stripped to low masses after mass transfer begins with a
partially hydrogen-depleted core. XTE~J1118+480, 1A~0620--00,
MAXI~J1305--704, and H~1705--250 occupy the 4--12.5~h region from which
such evolution may proceed, and are therefore useful candidate systems
for testing this pathway.

The orbital-position argument alone does not establish that any
individual observed binary will follow the transition branch.
Existing spectroscopy nevertheless provides partial but
heterogeneous evidence relevant to this interpretation. In
XTE~J1118+480, the ultraviolet accretion-flow spectrum shows prominent
N~{\sc v} emission while C~{\sc iv} and O~{\sc v} are undetected.
The anomalous N~{\sc v}/C~{\sc iv} line balance strongly indicates
CNO-processed transferred material
\citep{Haswell2002}. These emission lines do not constitute a
direct donor-photosphere abundance measurement, and their conversion
into a quantitative C/N abundance would require knowledge of the
geometry, density, and ionizing spectrum of the line-forming region.
In 1A~0620--00, near-infrared donor spectroscopy reveals strongly
weakened CO absorption, with the best-fitting synthetic donor spectra
requiring \([\mathrm{C/H}]\simeq-1.5\)
\citep{Froning2007}. This provides evidence for substantial
photospheric carbon depletion, consistent with exposure of
CNO-processed material, but nitrogen was not measured and hence no
direct photospheric C/N ratio is available. For MAXI~J1305--704, the
donor mass is substantially lower than that of a normal dwarf with its
observed early-K spectral type, indicating that the donor is evolved
\citep{MataSanchez2021}; this is an evolutionary rather than a chemical
diagnostic. Existing spectroscopy of H~1705--250 constrains its
late-type donor and dynamical properties
\citep{Filippenko1997,Harlaftis1997,Remillard1996}, but the cited
studies do not provide a quantitative C/N abundance. Direct
photospheric carbon and nitrogen measurements, particularly for
MAXI~J1305--704 and H~1705--250, would therefore provide valuable
independent tests of the proposed transition pathway, although their
faint quiescent donors make such observations challenging.
Quantitative comparison with the model maps requires atmosphere and
spectral-synthesis calculations.

No dynamically confirmed black-hole binary is presently known at the
predicted ultrashort periods of approximately 10--50~min.

\subsection[Lack of detections in predicted regions]{Lack of detections in predicted regions}
The calculations identify two regions of the
$P_{\rm orb}$--$M_2$ plane in which black-hole binaries may occur but no
secure example is presently known: detached black-hole--white-dwarf
binaries, and systems with $P_{\rm orb}\lesssim2$--3~h and
$M_2<0.1\,M_\odot$.

The detached binaries may eventually be found through radial-velocity or
astrometric studies of apparently isolated white dwarfs.
The absence of detected short-period, very-low-mass systems may
instead reflect either inefficient formation or observational selection
bias. In the stripped-donor BH sequences calculated here, systems that
reach this region generally have low secular mass-transfer rates. Within
the thermal--viscous disc-instability framework, a low supply rate can
produce long recurrence times between outbursts. At still lower rates,
a disc with a truncated inner region may remain globally stable on the
cold branch, producing only faint accretion emission rather than
detectable outbursts
\citep{FrankKingRaine2002,Lasota2001,HameuryLasota2016}. Such systems
would therefore be difficult to discover as X-ray transients and to
confirm dynamically in quiescence. Their present non-detection alone
cannot determine whether this evolutionary channel is intrinsically
rare or whether an existing Galactic population has largely escaped
detection.

Very faint X-ray transients provide a possible
observational connection to the low-transfer-rate sequences.
In the sample studied by \citet{DegenaarWijnands2009}, some accretors
were identified as neutron stars through type-I X-ray bursts,
while others had no secure identification and were also considered
as possible black-hole systems. The inferred long-term mean
accretion rates are low, but depend on the assumed accretor and
radiative efficiency. Component-mass and orbital-period measurements
are therefore needed to determine whether such sources occupy the
low-mass, short-period region predicted here.

The astrometric discovery of dormant black-hole
binaries demonstrates that X-ray selection misses viable
evolutionary products
\citep{ElBadry2023a,GaiaBH3,MuellerHorn2026}.
High-precision astrometry, spectroscopy and long-term photometric
monitoring can now test the long-period and low-transfer-rate
regions populated by the calculations.
These regions provide concrete target space for searches for
black-hole binaries outside the traditional transient X-ray
sample.

\subsection[Initial donor-star masses in black-hole binaries]{Initial donor-star masses in black-hole binaries}
Our sequences show that initially
intermediate-mass donors can be stripped into the low-mass regime,
as illustrated in the following subsection. The initial donor masses
therefore need not resemble the present-day masses of observed
BH--LMXB companions. These post-black-hole-formation calculations
do not, however, determine whether the low-mass initial companions
survive the preceding common-envelope phase.

\subsection[Effective temperatures of the donor stars]{Effective temperatures of the donor stars}
The cool, short-period locus in
Fig.~\ref{fig:Fig05} corresponds to the low-mass, short-period
region of the \(P_{\rm orb}\)--\(M_2\) diagrams in
Figs.~\ref{fig:Fig02} and~\ref{fig:Fig03}.
To characterise this locus, we consider the illustrative region
\(3\leq P_{\rm orb}/{\rm h}\leq12\) and
\(3000\leq T_{\rm eff}/{\rm K}\leq5000\).
Among actively transferring states with
\(\lvert\dot M_2\rvert\geq10^{-13}\,M_\odot\,{\rm yr}^{-1}\),
the current donor masses span approximately
\(0.13\)--\(0.84\,M_\odot\) in both samples.
Their initial donor masses span
\(0.70\)--\(2.06\,M_\odot\) in the GRID and
\(0.80\)--\(1.91\,M_\odot\) in the BPS sample.
This region therefore contains both initially low-mass donors
and initially intermediate-mass donors that have been stripped
by mass loss. For example, a GRID sequence beginning with
\(M_{2,\rm init}=2.06\,M_\odot\) reaches
\(M_2\simeq0.24\,M_\odot\) at
\(P_{\rm orb}\simeq6\,{\rm h}\) and
\(T_{\rm eff}\simeq4100\,{\rm K}\).

In Fig.~\ref{fig:Fig05}, the adopted effective
temperatures of known BH--LMXB donors are compared with the evolutionary
tracks in the \(T_{\rm eff}\)--\(P_{\rm orb}\) plane. Among the systems
with periods of approximately 4--16~h and inferred low-mass donors,
almost all lie on or near regions crossed by many tracks.
GRO~J0422+32 is the clearest exception: its adopted temperature of
approximately 3200~K lies about 880~K below the nearby track
concentration.

For GRO~J0422+32, the 3200~K value was fixed in
the light-curve modelling of \citet{Cherepashchuk2024}, who assumed an
M2~V donor; it is not an independent spectroscopic temperature
measurement. The M4--M5~V classification listed in Table~\ref{tab:BHList}
comes instead from the TiO spectroscopy of \citet{Webb2000}.
Other analyses favour an earlier type: \citet{GelinoHarrison2003}
obtained an M1 classification from the optical--infrared spectral
energy distribution and adopted 3900~K. 
These inconsistent temperature estimates need to be resolved before
GRO~J0422+32 can provide a decisive test of the tracks.

\citet{Justham2006} found a more systematic mismatch between observed
donor temperatures and calculations designed to address common-envelope
ejection with initially more massive Ap/Bp donors and associated magnetic
braking.  Such progenitors naturally remain hotter at comparable orbital
periods than the initially low-mass donors represented in much of the
present GRID. The substantially broader overlap in our
calculation therefore explains why the systematic offset found in that
restricted channel is not reproduced here. 

\subsection[Effects of irradiation]{Effects of irradiation}
The targeted irradiated calculations were used to assess how donor
heating modifies the mass-transfer history.  We adopt a nominal
irradiation efficiency $\eta_0=0.05$, representing the combined
effects of the incident accretion luminosity, disc shielding,
albedo, and redistribution over the donor surface. The effective
efficiency is reduced to $10^{-5}$ when the conditions in
Section~\ref{subsec:irradiation_method} are satisfied.  
Irradiation produces recurrent high- and low-transfer episodes, while the corresponding
irradiation-free histories are substantially smoother.

The full irradiated sample in
Fig.~\ref{fig:Fig08} shows the same broad division into converging
and expanding branches as the irradiation-free samples in
Figs~\ref{fig:Fig02} and~\ref{fig:Fig03}.
The three illustrative evolutions in
Section~\ref{sec:detailed_irradiation} show how the irradiation
cycles produce steps in donor mass and orbital period.
Their negligible-irradiation counterparts follow similar broad
trends and evolutionary timescales, with small differences
in donor mass and period at the same age.

Irradiation feedback on the donor can drive long-term variations
in the secular mass-transfer rate and alter the time spent at different
mass-transfer levels, as discussed for neutron-star binaries by, e.g.,
\citet{Pfahl2003}. These donor-response cycles are physically distinct
from the thermal--viscous outburst--quiescence cycles of the accretion
disc. Our calculations follow the donor's structural response to
irradiation on the adaptive stellar-evolution timestep, but they do not
evolve the disc instability. They therefore neither resolve individual
disc outbursts nor classify tracks as persistent or transient.
In the larger irradiated sample, the imposed reduction of
\(\eta_{\rm irr}\) on the lowest-mass short-period branch means
that the tracks do not test sustained strong irradiation down
to the smallest donor masses.
Future calculations should combine the
donor-response treatment with a separate time-dependent
irradiated-disc instability prescription, building on
\citet{Hameury1993} and subsequent work, and extend comparisons
at identical initial parameters across donor mass, initial
period, irradiation efficiency and disc duty cycle.

The long low-transfer intervals
($\gtrsim10^5$ yr) in the illustrative irradiation cycles
suggest that black-hole binaries close to Roche-lobe contact
could spend extended intervals as weakly accreting,
X-ray-faint systems. Such quiescent BH binaries should be searched for.
Finding systems with the predicted donor and orbital
properties would support the irradiation-cycle interpretation.

\subsection[Black-hole mass growth]{Black-hole mass growth}
The irradiation-free calculations also record the cumulative
black-hole mass growth.  The resulting map is strongly dependent on
donor mass, initial mass ratio, and sample coverage, with the largest
values concentrated in the high-donor-mass, near-unity-mass-ratio part
of the BPS-seeded sample.
These values are conditional on the idealized
accretion prescription adopted in the irradiation-free GRID
and BPS calculations, which adopts a nominal retained fraction
of unity for the transferred material.
The \(f_0=0.3\) prescription is not used in these
irradiation-free reference samples.
Because the accretor angular momentum is not evolved,
the mass-growth diagnostic must not be interpreted as a prediction of
natal or present-day black-hole spin.

\subsection[Potential LISA systems]{Potential LISA systems}
Only binaries that reach the shortest-period portions of our
evolutionary sequences are reasonable candidates for LISA detection.
For a source at \(d=1\,\mathrm{kpc}\), these systems have orbital periods
shorter than approximately \(6\)--\(12\,\mathrm{h}\), depending on the
sequence, and donor masses of approximately
\(0.02\)--\(0.12\,M_\odot\).
At \(d=1\,\mathrm{kpc}\), 2,642 tracks attain a
nominal four-year signal-to-noise ratio \(\rho\geq1\) at least once
during their calculated evolution, while 60 reach \(\rho\geq7\), our
adopted nominal detection threshold. At the maximum-\(h_{\rm c}\)
states displayed in Fig.~\ref{fig:Fig10}(b), 56 complete tracks satisfy
\(\rho\geq7\). The difference arises because a track's maximum
characteristic strain and maximum signal-to-noise ratio need not
occur at the same state.

These results demonstrate detectability at an assumed distance of
1~kpc and evolutionary reachability; they are not a LISA yield
prediction because the track ensembles carry neither binary-formation
weights nor a Galactic spatial and observational selection function.

\citet{DengLi2024} predicted that
\(\lesssim2.5\) per cent of their formation-weighted BH--LMXB
population evolves to orbital periods below 90 min.
That Galactic fraction is not directly comparable with the
2,624 tracks identified here, which constitute an unweighted
reachability count across the computed ensembles.
These compact black-hole binaries should nevertheless
be targeted by observational searches.

\subsection{Future work}
\label{subsec:future_work}
The present framework can be extended to test alternative
magnetic-braking prescriptions, including the convection and rotation
boosted (CARB) model of \citet{VanIvanova2019}, and different assumptions
for mass retention and the angular momentum carried away by escaping
material. Together with the proposed modelling of the donor's response to irradiation
and the time-dependent instability of the accretion disc,
these comparisons would help establish how the predicted evolutionary
paths and observable accretion states depend on the adopted physics.

A complementary step is to connect the evolutionary sequences to
updated models of binary formation and core collapse. For example,
\citet{Willcox2026} examine a remnant-mass prescription that produces a
bimodal black-hole birth-mass distribution and compare its consequences
with the chirp masses of merging binary black holes. Applying such
prescriptions to BH--LMXB progenitors could test how the black-hole
mass distribution and binary survival affect the systems entering the
X-ray binary phase. Combining formation rates, evolutionary lifetimes
and observational selection would then allow population predictions
beyond the evolutionary reachability mapped here.

\section{Conclusions}

We have followed the evolution of 78,795
irradiation-free black-hole binaries drawn from a regular GRID and a
BPS-seeded sample. Although their initial sampling differs, both recover
expanding giant-donor and converging short-period branches. The main
features of the evolutionary maps are therefore present in both of the
adopted initial distributions.

Mass transfer can strip initially intermediate-mass
donors into the low-mass regime and reproduce the cool companions
observed in many short-period BH--LMXBs. Some converging sequences
continue to substellar donor masses and ultrashort orbital periods,
where black-hole systems remain observationally unconfirmed. The
expanding branch provides a different outcome: stripped remnants in
wide, nearly circular orbits, providing a route to detached
black-hole--white-dwarf binaries.

The separate sample of 4,392 irradiated tracks
also exhibits the broad converging and expanding branches. Three
illustrative histories show recurrent irradiation-driven mass-transfer
episodes accompanied by steps in donor mass and orbital period.
Calculations with the same initial conditions and other input controls,
but negligible irradiation, follow similar secular trends and
evolutionary timescales. In these examples, the irradiation cycles
strongly change the timing of mass transfer while preserving the broad
pattern of orbital evolution. The selected events do not determine a
single duty cycle for the whole population.

The gravitational-wave analysis uses the
irradiation-free GRID and BPS samples. Some compact descendants reach
states above the adopted LISA detection threshold of \(\rho=7\) for
a four-year observation at a distance of 1~kpc. These calculations
identify potentially observable evolutionary stages. Predicting how
many such sources LISA will detect additionally requires their
formation rates, lifetimes and Galactic spatial distribution.

The calculations begin after the black hole has
formed. Whether all of the adopted initial binaries can be produced,
particularly those with low-mass companions that must survive a
preceding common-envelope phase, remains a separate formation problem.
During the subsequent black-hole--donor evolution, the appropriate
magnetic-braking prescription and the amount of mass and angular
momentum lost from the binary remain important uncertainties.
The irradiation calculations also require an improved connection
between the donor's mass-transfer rate and the accretion luminosity:
the present models follow the donor's response but do not evolve the
time-dependent accretion disc.

Observational selection must also be considered
when comparing these evolutionary paths with the known population.
Low-transfer-rate systems can be faint in X-rays, while detached
descendants are missed by searches that rely on accretion emission.
The calculated tracks show which states can be reached under the
adopted assumptions. Formation weights and survey selection functions
are needed to turn these results into predicted source numbers.

\section*{Acknowledgements}
We thank the anonymous referee for suggesting that we clarify our treatment of angular-momentum loss and discuss future extensions of this work. Computational resources were provided
thanks to support from Calcul Qu\'{e}bec and the Digital Research
Alliance of Canada (DRAC); TRUBA, through T\"{U}B\.{I}TAK
ULAKB\.{I}M; and the Cambridge Service for Data Driven Discovery
(CSD3), operated by the University of Cambridge Research Computing
Service. The DRAC systems (e.g., Fir) were also supported by the
Canada Foundation for Innovation, NanoQu\'{e}bec, RMGA, and the
Fonds de recherche du Qu\'{e}bec -- Nature et technologies (FRQNT).
The Cambridge infrastructure was provided by Dell EMC and Intel
using Tier-2 funding from the Engineering and Physical Sciences
Research Council (grant EP/T022159/1) and DiRAC funding from the
Science and Technology Facilities Council. This work was supported by the Scientific and Technological Research Council of T\"{u}rkiye (T\"{U}B\.{I}TAK) under project Nos
117F188 and 119F077. CAT and KY acknowledge
support from Churchill College, University of Cambridge, through
a Teaching Fellowship and an Overseas Fellowship, respectively.
LAN would like to thank the Natural Sciences and Engineering
Research Council (NSERC) of Canada for financial support and the
Digital Research Alliance of Canada for a 2026--27 Resource
Allocation Grant. H.-L.C. is partially supported by the National
Natural Science Foundation of China (grant Nos 12288102, 12333008
and 12422305), the National Key R\&D Program of China (grant Nos
2021YFA1600400/401 and 2021YFA1600403), the Yunnan Fundamental
Research Project (Nos 202401BC070007 and 202601CJ070008), the
International Centre of Supernovae (ICESUN), the Yunnan Key
Laboratory of Supernova Research (No. 202505AV340004), and the
Young Talent Project of the Yunnan Revitalization Talent
Support Program.

\section*{Data Availability}
The reduced evolutionary tracks and derived data products used for the
analysis and figures in this study are available from the corresponding
author upon reasonable request. The observed quantities used for
comparison were compiled from the published sources cited in the
corresponding tables.

\bibliographystyle{mnras}
\bibliography{BH_XRB} %

\FloatBarrier
\appendix
\section{Supplementary evolutionary diagnostics}
\label{sec:supplementary_diagnostics}

This Appendix presents supplementary donor-evolution and
surface-composition diagnostics that support the main analysis.
For the H--R-diagram diagnostic, we retain the full stored donor
histories, including pre-contact states, whereas the main
observational-plane maps begin at the onset of Roche-lobe overflow.

\begin{figure*}
\centering
\includegraphics[width=0.95\linewidth]{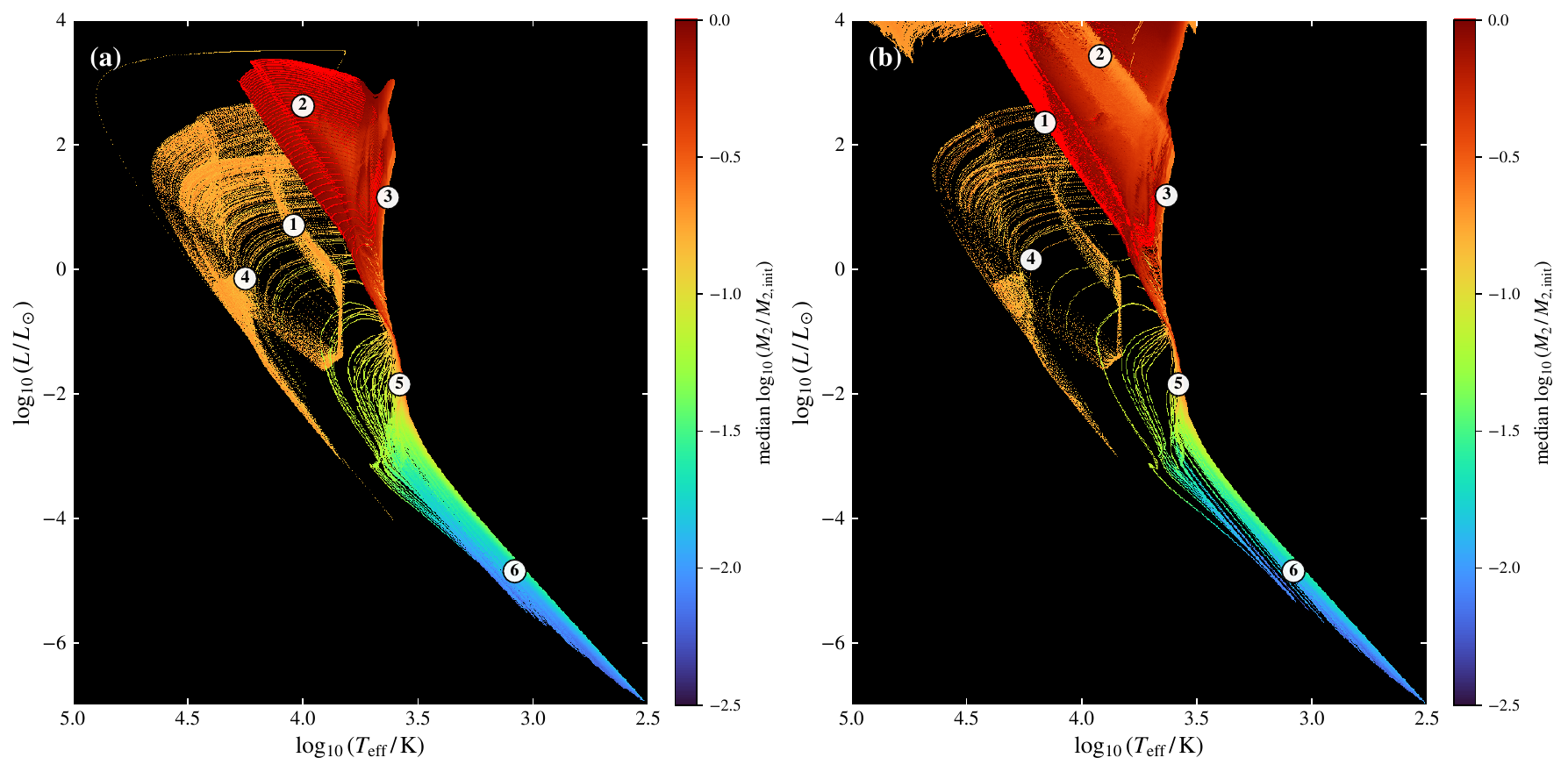}
\caption{Hertzsprung--Russell diagrams of the donor stars for
(a) the GRID and (b) the BPS-seeded calculations.
Small numbered circles mark six representative evolutionary
regimes. Their diagnostic definitions and physical interpretations are
given in Table~\ref{tab:hr_regions}. Colours show the pixel-wise median
\(\log_{10}(M_2/M_{2,\rm init})\). Values near zero identify donors that
retain nearly all their initial mass. Increasingly negative values indicate progressively stronger envelope stripping, \(-1\)
and \(-2\) corresponding to retaining 10 and 1 per cent, respectively.
The same colour limits,
\(-2.5\leq\log_{10}(M_2/M_{2,\rm init})\leq0\), are used in both
panels. The numbered regions are defined independently by the
diagnostic selections in Table~\ref{tab:hr_regions}.}
\label{fig:A1}
\end{figure*}

To interpret the branch morphology in
Fig.~\ref{fig:A1}, we selected donor records using their instantaneous
\(P_{\rm orb}\), \(M_2\), \(M_2/M_{2,\rm init}\), central hydrogen abundance \(X_{\rm c}\), and mass-transfer rate, and mapped these selections back
onto the H--R plane. Table~\ref{tab:hr_regions} summarizes the resulting
physical regimes. The selections are representative and non-exclusive;
the numbered labels identify characteristic evolutionary branches
rather than providing an exhaustive classification of every stored
model state. The colour progression makes the degree of stripping directly
visible: donor records along the converging branch and the
very-low-mass tail retain progressively smaller fractions of their
initial masses than records near the weakly stripped upper loci.

\begin{table*}
\caption{Physical interpretation of the numbered donor
H--R-diagram regions in Fig.~\ref{fig:A1}. The selections use
instantaneous model properties and are representative, non-exclusive
diagnostic classifications, not population weights.}
\label{tab:hr_regions}
\centering
\small
\begin{tabular}{@{}c
  >{\raggedright\arraybackslash}p{0.17\textwidth}
  >{\raggedright\arraybackslash}p{0.43\textwidth}
  >{\raggedright\arraybackslash}p{0.26\textwidth}@{}}
\toprule
Region & H--R morphology & Physical interpretation &
Diagnostic model selection \\
\midrule

1 &
Hot diagonal locus &
Hydrogen-rich donors that remain close to their initial mass; these
records trace pre-contact or early-RLOF evolution before substantial
envelope stripping. &
\(M_2/M_{2,\rm init}\geq0.9\) and \(X_{\rm c}>0.05\) \\
&  &  & \\ 
2 &
Upper hot/luminous fan &
A mixture of nuclear-expanded donors and donors undergoing rapid
envelope stripping. The H--R position alone does not distinguish
thermal readjustment from rapid nuclear expansion. &
\(P_{\rm orb}\geq10\) h and either
[\(X_{\rm c}\leq0.05\), \(M_2/M_{2,\rm init}\geq0.8\)] or
[\(M_2/M_{2,\rm init}<0.8\), \(M_2\geq1\,M_\odot\),
\(|\dot M_2|\geq10^{-8}\,M_\odot\,{\rm yr}^{-1}\)] \\
&  &  & \\
3 &
Cool, luminous near-vertical plume &
Hydrogen-exhausted donors with stripped envelopes that remain in active
mass transfer on the long-period expanding branch. &
\(P_{\rm orb}\geq10\) h, \(0.1\leq M_2/M_\odot<1\),
\(M_2/M_{2,\rm init}<0.7\), \(X_{\rm c}\leq0.05\), and
\(|\dot M_2|\geq10^{-12}\,M_\odot\,{\rm yr}^{-1}\) \\
&  &  & \\
4 &
Hot, lower-luminosity loops &
Long-period, strongly stripped donors in detached or nearly detached
contraction; their locations are proto-He-WD-like, but the H--R diagram
alone does not establish a final remnant class. &
\(P_{\rm orb}\geq10\) h, \(M_2<0.35\,M_\odot\),
\(M_2/M_{2,\rm init}<0.3\), \(X_{\rm c}\leq0.05\), and
\(|\dot M_2|<10^{-12}\,M_\odot\,{\rm yr}^{-1}\) \\
&  &  & \\
5 &
Faint hook near \(\log T_{\rm eff}\simeq3.6\) &
Actively transferring low-mass donors that predominantly occupy
the angular-momentum-loss-driven converging BH--LMXB sequence. &
\(P_{\rm orb}<10\) h, \(0.1\leq M_2/M_\odot<1\),
\(M_2/M_{2,\rm init}<0.9\), and
\(|\dot M_2|\geq10^{-12}\,M_\odot\,{\rm yr}^{-1}\) \\

&  &  & \\
6 &
Cool diagonal tail to very low luminosity &
Very-low-mass and substellar donors near or beyond the period minimum.
Numerical stopping points along this tail are not physical endpoints. &
\(P_{\rm orb}\leq3\) h and \(M_2<0.1\,M_\odot\) \\
\bottomrule
\end{tabular}
\end{table*}

\begin{figure*}
\centering
\includegraphics[width=0.95\linewidth]{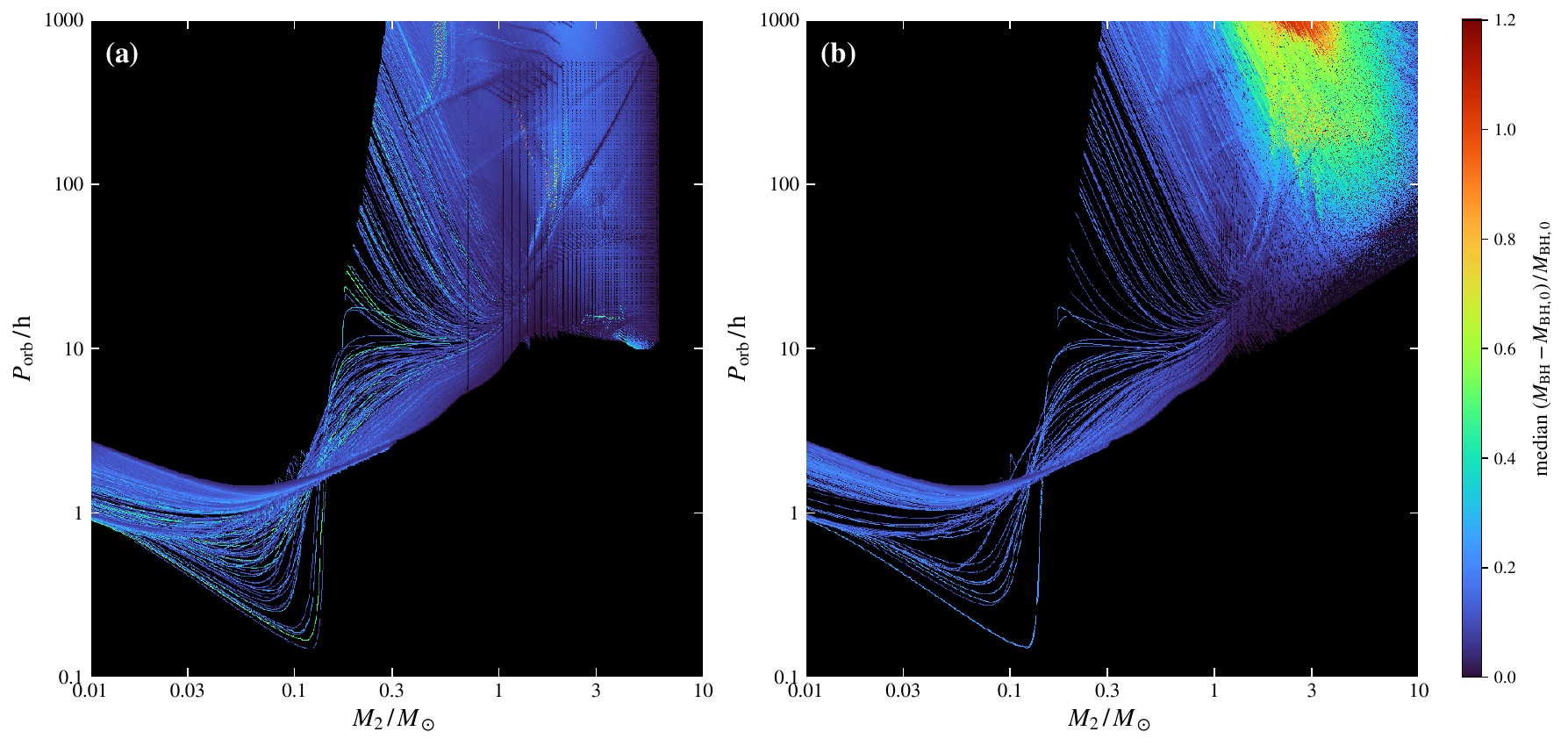}
\caption{Orbital-period--donor-mass plane colour-coded by the
pixel-wise median fractional black-hole mass growth,
\(g_{\rm BH}\), for
(a) the irradiation-free GRID and (b) the irradiation-free BPS-seeded
calculations. Here \(M_{\rm BH,0}\) is the black-hole mass in the first
valid stored state of each sequence. Both panels use the same colour
scale. The plotted \(g_{\rm BH}\) is a cumulative mass increase
normalized by \(M_{\rm BH,0}\), not the instantaneous retained fraction
\(f_{\rm acc}\), and therefore is not bounded above by 0.3. These legacy
irradiation-free calculations nominally set the pre-limit retained
fraction to unity. The \(\beta_{\rm MT}=0.7\), \(f_0=0.3\) prescription
used in the separate targeted irradiated production grids does not
enter this figure. The stronger signal in the upper-right part of the
BPS panel reflects its larger initial donor masses and near-unity
initial mass ratios under this idealized accretion prescription. The
figure is therefore a sampling- and prescription-dependent diagnostic
of the stored model mass histories, not a prediction of black-hole spin
or of a population-weighted Galactic distribution.}
\label{fig:A2}
\end{figure*}

The irradiation-free calculations also record the growth of the
black-hole mass. We define
\[
g_{\rm BH}\equiv
\frac{M_{\rm BH}-M_{\rm BH,0}}{M_{\rm BH,0}},
\qquad
M_{\rm BH,0}\equiv M_{\rm BH}(t_{\rm first}).
\]
Here \(t_{\rm first}\) denotes the evolutionary time of the
first valid stored state included in the analysis for a given track.
Here \(g_{\rm BH}\) is the cumulative black-hole mass increase normalized
by the black hole's mass in the first valid stored state. It is not the
instantaneous retained fraction \(f_{\rm acc}\). The legacy custom
implementation used for the irradiation-free GRID and BPS calculations
nominally sets the pre-limit retained fraction to unity. 
The \(\beta_{\rm MT}=0.7\),
\(f_0=0.3\) prescription is not used in the irradiation-free
reference samples shown in Fig.~\ref{fig:A2}.
Consequently, \(g_{\rm BH}\) is not bounded above by
0.3 and can approach or exceed unity when the accumulated mass gain is
comparable to the initial black-hole mass. We show this model quantity
directly rather than converting it into a spin estimate, because the
calculations do not evolve the angular momentum of the accretor.
Large values of $g_{\rm BH}=(M_{\rm BH}-M_{\rm BH,0})/M_{\rm BH,0}$  are therefore conditional on
the idealized legacy accretion prescription and are not predictions of
natal or present-day black-hole spin.

\begin{figure*}
\centering
\includegraphics[width=0.8\linewidth]{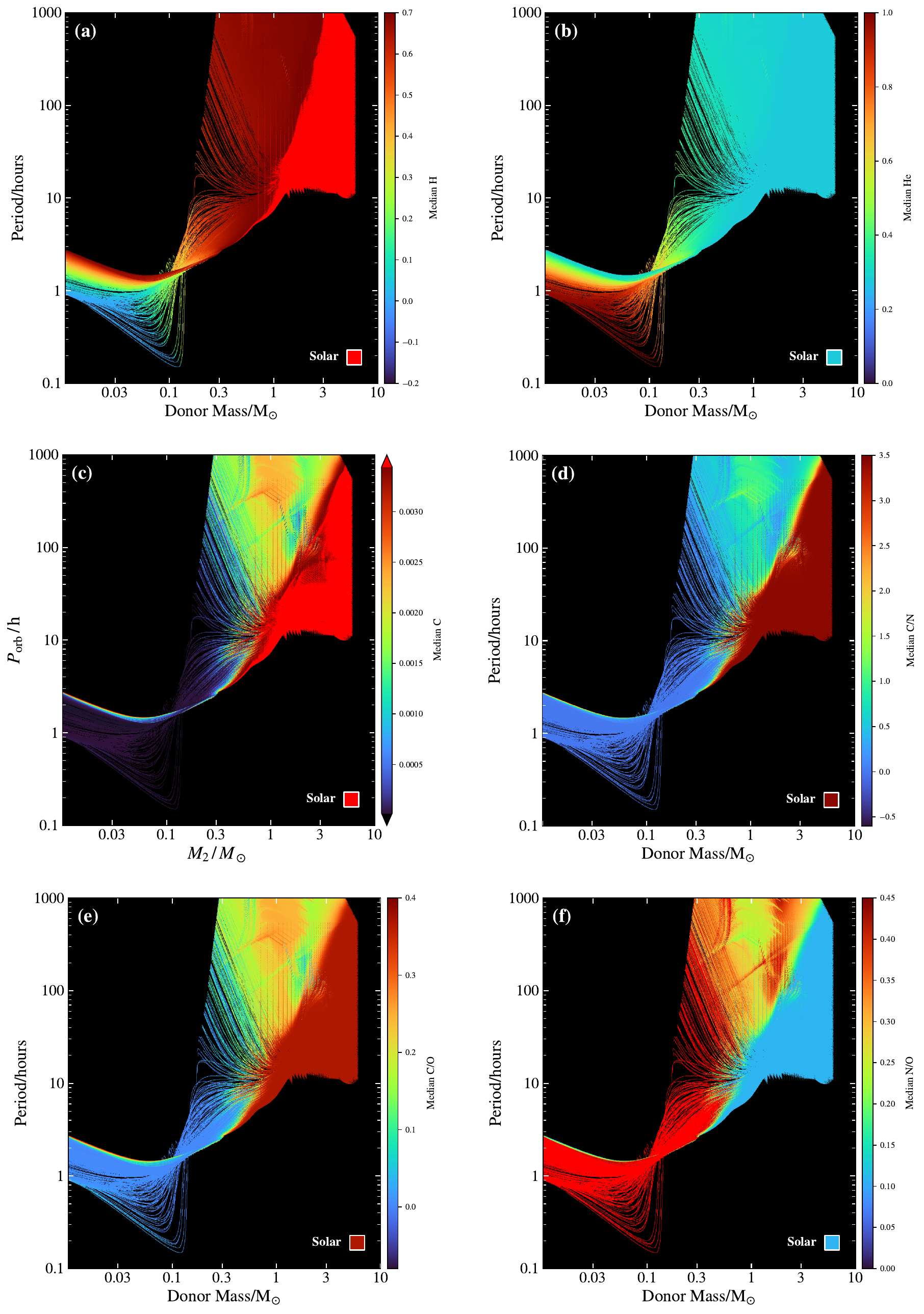}
\caption{Donor surface-abundance diagnostics in the
orbital-period--donor-mass plane for the irradiation-free GRID calculations,
(a) \(X_{\rm H}\), (b) \(Y_{\rm He}\), (c) \(X_{\rm C}\), (d) C/N,
(e) C/O and (f) N/O. Colours give the pixel-wise median of the
corresponding surface quantity in each bin. All quantities are shown on linear scales.
The coloured square labelled ``Solar'' in the lower-right corner
of each panel gives the corresponding abundance in the adopted initial
solar-composition mixture,
\(X_{\rm H,\odot}=0.7000\),
\(Y_{\rm He,\odot}=0.27997\),
\(X_{\rm C,\odot}=3.4421\times10^{-3}\),
\(({\rm C/N})_\odot=3.4128\),
\(({\rm C/O})_\odot=0.36764\) and
\(({\rm N/O})_\odot=0.10772\).
Quantitative comparison with observed spectra would require
stellar-atmosphere and spectral-synthesis calculations.}
\label{fig:A3}
\par\raggedright
\bsp
\label{lastpage}
\end{figure*}

\end{document}